\documentclass[aps,prd,twocolumn,superscriptaddress,preprintnumbers,floatfix,longbibliography]{revtex4-1}

\usepackage{dcolumn}% Align table columns on decimal point
\usepackage{bm}% bold math
\usepackage{graphicx}
\usepackage{amsmath}
\usepackage[caption=false]{subfig}
\usepackage{siunitx}
\usepackage{placeins}
\usepackage{color}
\usepackage{standalone}
\usepackage{dcolumn}
\usepackage{bm}
\usepackage{microtype}
\usepackage{etoolbox}
\usepackage{amssymb}
\usepackage{mathrsfs}
\usepackage{accents}
\usepackage[normalem]{ulem}
\usepackage[dvipsnames]{xcolor}
\usepackage[colorlinks,urlcolor=NavyBlue,citecolor=NavyBlue,linkcolor=NavyBlue,pdfusetitle]{hyperref}
\usepackage{gensymb}
\AtBeginDocument{\usepackage{booktabs}}
\graphicspath{{./figs/}}
\usepackage[colorlinks,urlcolor=NavyBlue,citecolor=NavyBlue,linkcolor=NavyBlue,pdfusetitle]{hyperref}

\newcommand{\msun}{M_{\odot}}
\newcommand{\ra}{{\alpha_\star}}
\newcommand{\dec}{{\delta_\star}}
\newcommand{\fdot}{\dot{f_0}}
\newcommand{\Tobs}{T_{\rm obs}}
\newcommand{\Tcoh}{T_{\rm coh}}
\newcommand{\dcc}{LIGO--P1900274}

\newcommand{\heading}[1]{\section{#1}}
\newcommand{\subheading}[1]{\subsection{#1}}

\begin{document}
	
%\preprint{APS/123-QED}

\title{Search for ultralight bosons in Cygnus X-1 with Advanced LIGO}% Force line breaks with \\

\author{Ling Sun}
\email[]{ling.sun@anu.edu.au}
\affiliation{LIGO Laboratory, California Institute of Technology, Pasadena, California 91125, USA}
\affiliation{OzGrav-ANU, Centre for Gravitational Astrophysics, College of Science, The Australian National University, ACT 2601, Australia}

\author{Richard Brito}
\email[]{richard.brito@roma1.infn.it}
\affiliation{Dipartimento di Fisica, ``Sapienza'' Universit\`a di Roma \& Sezione INFN Roma1, Piazzale Aldo Moro 5, 00185, Roma, Italy}

\author{Maximiliano Isi}
\email[]{maxisi@mit.edu}
\thanks{NHFP Einstein fellow}
\affiliation{
	LIGO Laboratory, Massachusetts Institute of Technology, Cambridge, Massachusetts 02139, USA}%

\date{\today}% It is always \today, today,

\begin{abstract}
	
Ultralight scalars, if they exist as theorized, could form clouds around rapidly rotating black holes. Such clouds are expected to emit continuous, quasimonochromatic gravitational waves that could be detected by LIGO and Virgo. Here we present results of a directed search for such signals from the Cygnus X-1 binary, using data from Advanced LIGO's second observing run. We find no evidence of gravitational waves in the 250--750\,Hz band. Without incorporating existing measurements of the Cygnus X-1 black hole spin, our results disfavor boson masses in $6.4 \leq \mu/(10^{-13}\,{\rm eV}) \leq 8.0$, assuming that the black hole was born $5 \times 10^6$ years ago with a nearly-extremal spin. We then focus on a string axiverse scenario, in which self-interactions enable a cloud for high black-hole spins consistent with measurements for Cygnus X-1.
In that model, we constrain the boson masses in $9.6 \leq \mu/(10^{-13}\,{\rm eV}) \leq 15.5$ for a decay constant $f_a\sim 10^{15}$\,GeV.
Future application of our methods to other sources will yield improved constraints.
\end{abstract}

\pacs{Valid PACS appear here}% PACS, the Physics and Astronomy
\maketitle

%\tableofcontents

\heading{Introduction}
Ultralight scalar (spin 0) or vector (spin 1) boson particles have been theorized under several frameworks to solve problems in particle physics, high-energy theory and cosmology \cite{Peccei1977,Peccei1977PhRvD,Weinberg1978,Arvanitaki2010,Goodsell2009,Jaeckel:2010ni,Essig:2013lka,Hui:2016ltb}.
If such a new fundamental field exists, its occupancy number should superradiantly grow around fast-spinning black holes (BHs).
This occurs when $\omega_\mu / m < \Omega_{\rm BH}$, where $\omega_\mu=\mu/\hbar$ is the characteristic angular frequency of a boson with rest energy $\mu$, $m$ is the boson azimuthal quantum number with respect to the BH's rotation axis, and $\Omega_{\rm BH}$ is the angular speed of the outer horizon.
The superradiant instability is maximized when the Compton wavelength of the particle is comparable to the characteristic length of the BH, meaning $hc/\mu \sim GM/c^2$ for BH mass $M$.
If these conditions are satisfied, the number of ultralight bosons around the BH grows exponentially, forming a macroscopic cloud holding up to ${\sim}10\%$ of the BH mass.
This cloud can have a long lifetime, during which it generates continuous, quasi-monochromatic gravitational waves (GWs) \cite{Arvanitaki2011,Yoshino2014,Yoshino2015,Arvanitaki2015,Arvanitaki2017,Brito2017-letter,Brito2017,Baryakhtar2017}. 

By detecting such signals, ground-based instruments like Advanced LIGO (aLIGO) \cite{LIGO2014} and Virgo \cite{Virgo2014} could probe bosons with masses \mbox{$\sim 10^{-14}$--$10^{-11}$\,eV}, which are largely inaccessible to other experiments \cite{Arvanitaki2015,Brito2017-letter,Brito2017,Isi2019}. 
A search for a stochastic GW background from boson clouds in the first aLIGO observing run excluded a mass range of \mbox{$2.0 \leq \mu/(10^{-13}\,{\rm eV}) \leq 3.8$} at 95\% credibility, under optimistic assumptions about BH populations \cite{Tsukada2019}. 
Methods have been developed to search for continuous GWs from individual clouds, with and without restrictions to specific sky locations \cite{DAntonio2018,Isi2019}. 
Constraints on the boson mass ($\mu \sim 10^{-13}$\,eV) have been suggested using pre-existing strain upper limits for continuous GWs obtained from undirected searches \cite{Dergachev2019,Palomba2019}.
Like for the stochastic background, such constraints are contingent on BH populations.
\citet{Isi2019} modeled the signal waveforms for individual sources with a known sky location, and demonstrated the suitability of a specific search algorithm based on a hidden Markov model (HMM) to efficiently search for such signals \cite{Suvorova2016,ScoX1ViterbiO1,Sun2018,Isi2019}.
Two primary types of sources are of interests for such directed searches: remnants from compact binary coalescences (CBCs) \cite{LVC-catalog}, and known BHs in X-ray binaries \cite{Remillard:2006fc,Yoshino2015,Middleton2016}. 
Detection prospects for CBC remnants are hurt by their typically large luminosity distances, most likely demanding third-generation detectors \cite{Isi2019,Ghosh:2018gaw,Hild:2010id,Sathyaprakash:2012jk,Punturo:2010zz,Abbott2017-nextGen-CE}.
On the other hand, X-ray binaries have the advantages of being much closer and better localized, hence potentially lying within the sensitive range of existing detectors.
Constraints on the mass of axion-like particles have been suggested from spin measurements of BHs in X-ray binaries, roughly disfavoring a range of \mbox{$6\times 10^{-13} \leq \mu/{\rm eV}\leq 10^{-11}$} \cite{Arvanitaki2015,Cardoso2018,Stott:2018opm}.
For the constraints above, it is implicitly assumed that the boson does not self-interact significantly.
This would be the case, e.g., for a quantum-chromodynamics (QCD) axion with decay constant $f_a$ above the grand unification (GUT) scale \footnote{Since the QCD axion mass largely depends on $f_a$ of the Peccei-Quinn symmetry, e.g., $\mu \approx 6\times10^{-10}\,{\rm eV}(10^{16}\,{\rm GeV}/f_a)$ \cite{Arvanitaki2010}, the BHs observable by aLIGO correspond to axions with $f_a$ between the GUT and Planck scales.}.
However, nonlinear self-interaction could be significant in other proposals, like string axions, if $f_a$ were smaller than the GUT scale \cite{Arvanitaki2010,Arvanitaki2015,Yoshino2012,Yoshino:2015nsa,Yoshino2015}. Constraints on boson mass and decay constant can be studied by taking into consideration the nonlinear self-interaction in those scenarios, e.g., string axiverse.

In this article, we present results from a search for GWs from ultralight scalars in the X-ray binary Cygnus X-1 (Cyg X-1), using data from aLIGO's second observing run (O2) \cite{GWOSC,Vallisneri:2014vxa}.
Directed GW searches for sources within X-ray binaries are challenging because of the Doppler modulation induced on the GW by the binary motion.
The intrinsic, quasimonochromatic signal is shifted to lower and higher frequencies, resulting in a comb of orbital sidebands when analyzed in the frequency domain.
A matched filter is needed to collect the distributed signal power from the sidebands, whose width depends on the intrinsic signal frequency, the BH's projected semimajor axis, and the binary's orbital period.
% \red{Unfortunately, the orbital parameters are not well measured electromagnetically for most of the interesting X-ray binaries, and the sidebands generally spread too broadly ($\gtrsim 1$\,Hz) for existing search methods.}
%\mi{I would remove the above sentence---it doesn't add much and sounds too negative. You can just say below ``orbital parameters for other sources are generally not known to high-enough precision''.}
Cyg X-1 is one of the most interesting sources, with or without considering boson self-interactions, because of its relatively high BH mass ($14.8 M_\odot$), close proximity to Earth ($1.86$\,kpc), and relatively well measured orbital parameters.
In the search, we take advantage of the frequency-domain matched filter of Refs.\cite{Suvorova2016,ScoX1ViterbiO1} to account for the Doppler modulation.

Unlike for CBC remnants, there is large uncertainty about the age and spin of most BHs in X-ray binaries \cite{Reynolds:2013qqa,McClintock:2013vwa}.
Moreover, the impact of accretion from the companion is not perfectly understood \cite{Arvanitaki2015,Baryakhtar2017}.
Boson constraints derived from X-ray binaries, including those presented here, thus require assuming that a cloud would be sufficiently long-lasting to be present at the time of observation and that accretion does not substantially affect its formation.
When applicable, our constraints factor in the estimated age of Cyg X-1 (4.8--7.6 million years \cite{Gou2011,Wong2012}) in computing expected strain amplitudes.
Another potential issue comes from the spin of the BH in Cyg X-1, which some measurements indicate would be too high ($\geq 0.95$) to support a boson cloud in the simplest scenarios \cite{Gou2011,Axelsson2011,Walton2016}.
However, there seems to be disagreement in the literature about the BH spin, with some estimates favoring lower values \cite{Miller2009,Kawano2017,Krawczynski2018}.
We interpret our results under models with and without the assumption of high spin in Cyg X-1.

%Below, we briefly describe the analysis setup, and present the results of a directed search for GW signals from ultralight bosons in Cyg X-1 in the second observing run (O2) of aLIGO.
In the absence of a detection, we disfavor scalar masses in \mbox{$6.4 \leq \mu/(10^{-13}\,{\rm eV}) \leq 8.0$}, assuming that the BH has an age of $5 \times 10^6$\,yr and that it was born with a nearly-extremal spin but has an unknown post-superradiance spin.
Assuming a high post-superradiance spin, we also consider a specific scenario of string axiverse \cite{Arvanitaki2010,Arvanitaki2015,Yoshino2012,Yoshino:2015nsa,Yoshino2015}, with a decay constant $f_a \sim 10^{15}$\,GeV excluding the mass range \mbox{$9.6 \leq \mu/(10^{-13}\,{\rm eV}) \leq 15.5$}.
Below, we briefly describe the analysis setup, outline the results and their limitations,
and close with future prospects.

% \vspace{\baselineskip}
\heading{Method and setup}
The semi-coherent search is based on a HMM tracking scheme combined with a frequency domain matched filter, Bessel-weighted $\mathcal{F}$-statistic \cite{Isi2019,Suvorova2016,ScoX1ViterbiO1} (see Appendix) \footnote{An improved HMM method, which tracks the binary orbital phase and sums the signal power in orbital sidebands coherently, proves to be more sensitive \cite{Suvorova2017}. However, the improved method depends on the measurement of the time of passage through the ascending node, which is not available for Cyg X-1. Hence we do not apply the orbital phase tracking in this analysis.}. 
This efficient search strategy, which achieves the same sensitivity as other stack-slide-based semi-coherent algorithms, surmounts some of the computing challenges arising when the orbital parameters are not perfectly measured, and allows for uncertainties in the theoretical prediction of the signal model, e.g., cloud perturbations due to the astrophysical environment.
The total observing time $\Tobs$ is divided into shorter intervals with duration $\Tcoh$, over which the signal power is collected coherently.
The segments are labeled by discrete time steps $t_k$, for $k\in[0,N_T]$ and \mbox{$N_T=\Tobs/\Tcoh-1$}.
Over each interval $[t_k,t_k+\Tcoh]$, the intrinsic GW signal frequency $f_0$ is assumed to be monochromatic, remaining in one discretized frequency bin of width $\Delta f = 1/(2 \Tcoh)$. The signal power in each bin is estimated using a matched filter that accounts for Doppler modulation due to the motion of the source within the binary. The central value of bin $i$ is denoted $f_i$ with $i\in[1,N_Q]$, where $N_Q$ is the total number of frequency bins. We adopt the signal model described in Ref.~\cite{Isi2019} and assume that $f_0$ can evolve for at most one bin from $t_k$ to $t_{k+1}$. The HMM is solved by the classic Viterbi algorithm \cite{Viterbi1967}, returning the optimal path of signal frequency evolution $f_0^*(t_k)$ for $0\leq k \leq N_T$.

\begin{table}
	\begin{ruledtabular}
		\caption{Cygnus X-1 parameters.}
		\label{tab:cygX1-paras}
		\setlength{\tabcolsep}{0.2pt}
		\begin{tabular}{lccc}
			Parameter  & Symbol& Value & Ref.\\
			\midrule
			Black hole mass ($\msun$) & $M$& $14.8\pm1.0$ & \cite{Orosz2011}\\
			Mass ratio & $q$ & $1.29\pm0.15$ & \cite{Casares2014}\\
			Spin & $\chi$& $\geq 0.95$& \cite{Gou2011}\\
			Age (yrs) & $t_{\rm age}$ & $[4.8, 7.6]\times 10^6$ & \cite{Gou2011,Wong2012}\\
			Right ascension & $\ra$& $19^{\rm h}58^{\rm m}22^{\rm s}$ & \cite{Reid2011}\\
			Declination  & $\dec$ & $35^{\circ}12'0.6''$ & \cite{Reid2011}\\
			Inclination (deg) & $\iota$& $27.1\pm0.8$&\cite{Orosz2011}\\
			Distance (kpc) & $d$& $1.86\pm 0.12$ &\cite{Casares2014}\\
			Orbital period (days) & $P$& $5.599829\pm 0.000016$ &\cite{Orosz2011}\\
			Proj.\ semimajor axis (l-s) & $a_0$& $25.56^{+3.15}_{-3.11}$&\cite{Orosz2011}\\
		\end{tabular}
	\end{ruledtabular}
\end{table}

Based on the source parameters measured electromagnetically (given in Table \ref{tab:cygX1-paras}), we search a frequency band of 250--750\,Hz.
The expected signal strain $h_0$ would be too weak ($\lesssim 8 \times 10^{-26}$) to be detectable below 250\,Hz~\footnote{Note that throughout this paper, the GW strain $h_0$ differs from the numerically estimated strain in Eqn.~(28) of Ref.~\cite{Isi2019} due to different conventions. The strain in Eqn.~(28) of \cite{Isi2019} needs to be multiplied by a factor of $\sqrt{5/(4\pi)}$ to be directly comparable to the $h_0$ in this paper \cite{Isi2019-erratum,Sun2020-erratum}.}, and the orbital sidebands are too wide ($\gtrsim 0.5$\,Hz) to achieve the desired sensitivity above 750\,Hz~\cite{Isi2019}.
Given the source parameters and frequency band, we assume that the first time derivative of the GW frequency is $\fdot \sim 10^{-14}$--$10^{-13}$\,Hz/s \cite{Arvanitaki2015,Isi2019}, and hence select $\Tcoh=10$\,d ($\Delta f = 5.8 \times 10^{-7}$\,Hz) to cover a $\fdot$ range of $0 \leq \fdot \leq 6.7\times10^{-13}$\,Hz/s. Besides this, accretion could result in a small frequency variation due to secular changes in the BH parameters. However, since the typical accretion timescale is $t_{\rm acc} \sim 4.5\times 10^{7}$\,yrs at the Eddington rate, the frequency variation due to accretion for Cyg X-1 should be at most $\fdot \sim -8\times 10^{-16}(\alpha/0.1)^3(14.8 M_{\odot}/M)(4.5\times 10^{7}\,{\rm yrs}/t_{\rm acc})$\,Hz/s. Because this is, in general, much smaller than the variation due to the cloud dissipation \cite{Arvanitaki2015,Isi2019}, we neglect this effect.
We search over several values of the light-travel time across the projected semi-major axis of the orbit, in the range $22.45 \leq a_0/$(l-s)\,$\leq 28.71$ with bin size 0.05\,l-s.
This covers the uncertainty implied by the BH mass ($M$), companion mass, and inclination angle ($\iota$) measurements.

The search is parallelized into 1-Hz sub-bands. The detection score $S$ is defined, such that the log likelihood of the optimal Viterbi path equals the mean log likelihood of all paths plus $S$ standard deviations in each sub-band. A detection threshold $S_{\rm th} = 6.22$ for 1\% false alarm probability is determined through Monte-Carlo simulations, such that searching data sets containing pure noise yields 1\% of positive detections with $S>S_{\rm th}$. 

\begin{figure}[!tbh]
	\centering
	\includegraphics[width=\columnwidth]{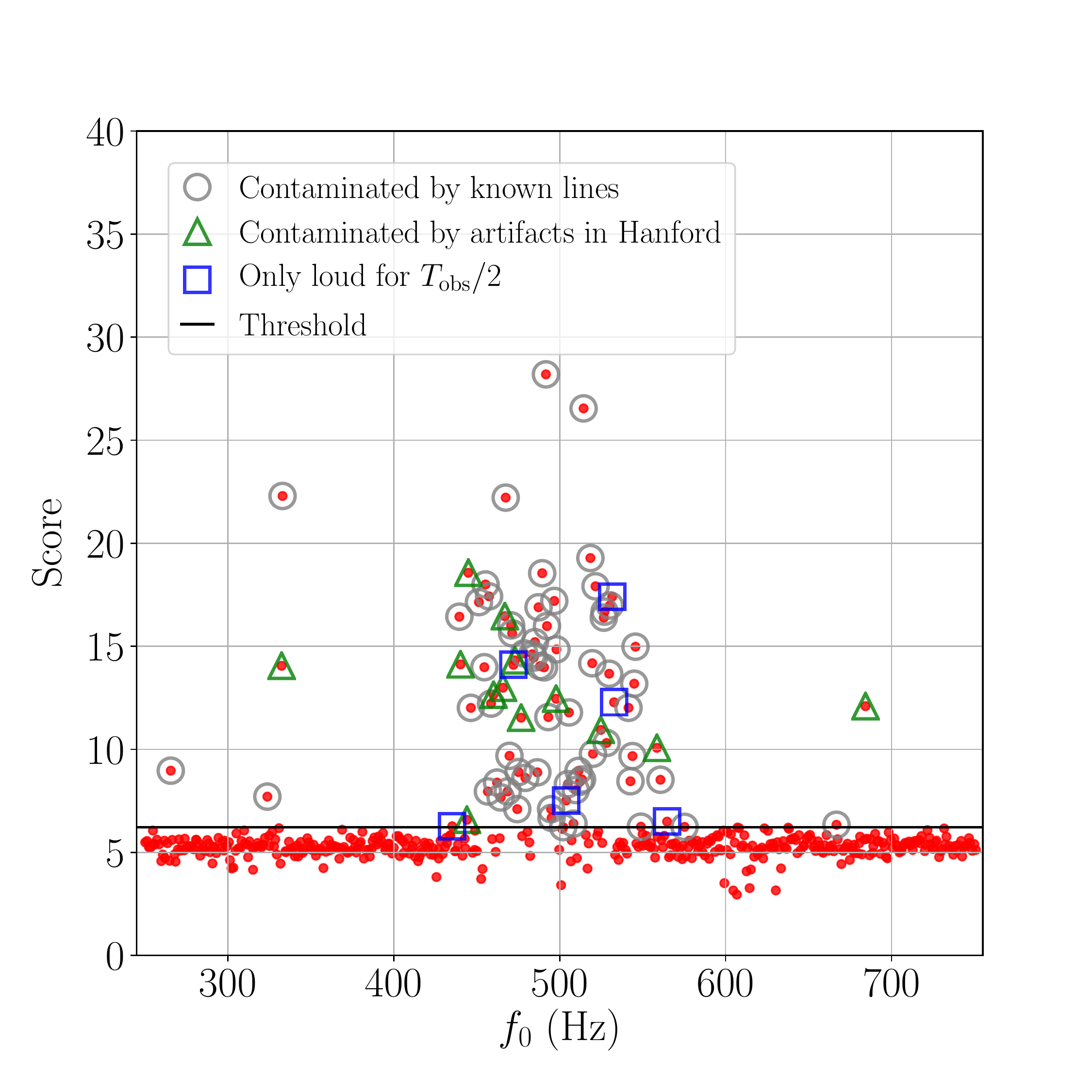}
	\caption[]{Detection score $S$ in each 1-Hz sub-band as a function of $f_0$. Red dots above the black line (1\% false alarm probability threshold $S_{\rm th} = 6.22$) are the first-pass candidates. Red dots marked by grey circles are vetoed due to contamination by known instrumental lines. Candidates marked by green triangles are vetoed because their scores are increased when analyzing Hanford only rather than the two detectors combined but below-threshold when analyzing Livingston only. Candidates marked by blue squares are vetoed because their scores are increased in one half of $\Tobs$ but below-threshold in the other half. No candidate survives all vetoes.}
	\label{fig:candidates}
\end{figure}

\begin{figure}[!tbh]
	\centering
	\includegraphics[width=\columnwidth]{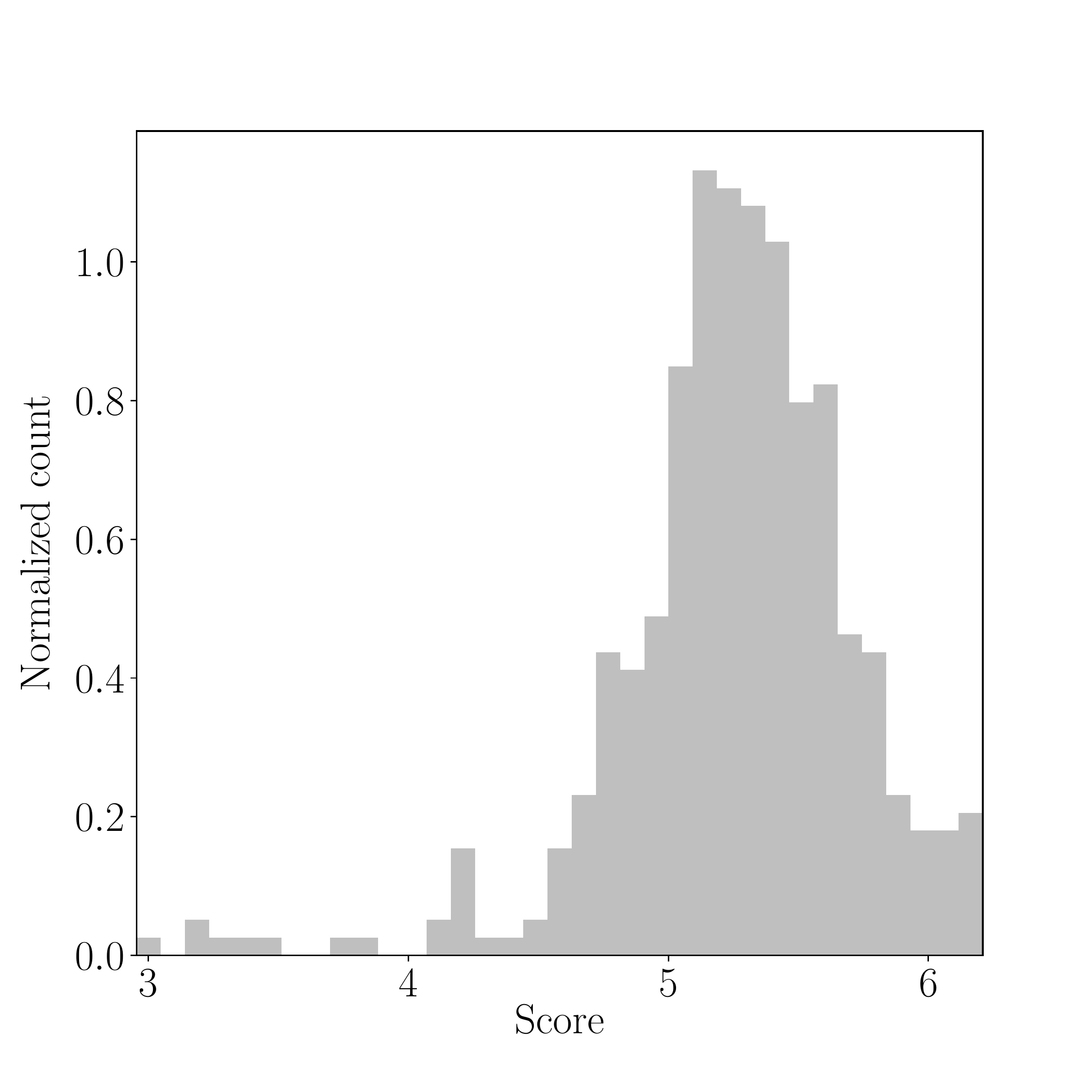}
	\caption[]{Distribution of the below-threshold scores from all 1-Hz sub-bands searched. The right edge of the plot indicates the 1\% false alarm probability threshold $S_{\rm th} = 6.22$.}
	\label{fig:hist}
\end{figure}

% \vspace{\baselineskip}
\heading{Results}
We analyzed aLIGO O2 data extending from 4 January 2017 to 25 August 2017 UTC (GPS time 1167545066 to 1187733592)\cite{GWOSC,Vallisneri:2014vxa} \footnote{The data collected in December 2016 is not used, because there is an end-of-year break starting on 22 December 2016 and the data quality before that (at the beginning of O2) is not optimal.}.
The search results are recorded in Fig.~\ref{fig:candidates}. Each red dot stands for the detection score $S$ obtained in a 1-Hz sub-band. The black line indicates $S_{\rm th} = 6.22$.
We claim a detection if a candidate with $S>S_{\rm th}$ passes a well-defined hierarchy of vetoes and is not identified as originating from an instrumental artifact.
We follow up the first-pass candidates found with $S>S_{\rm th}$ (83 in total), finding that none survives a three-stage veto procedure.
First, we find that 64 candidates overlap known instrumental lines (grey circles).
Second, we eliminate an additional 13 candidates because their significance is higher when analyzing Hanford only rather than the two detectors combined, while doing the same for Livingston yields $S<S_{\rm th}$.
This indicates contamination from noise artifacts in Hanford (green triangles).
Third, we veto the remaining 6 candidates because their significance is increased when searching one half of $\Tobs$, while the other half yields $S<S_{\rm th}$ (blue squares).
A full description of the veto procedure can be found in Ref.~\cite{ScoX1ViterbiO1}. The distribution of all the scores obtained in the sub-bands without contamination is shown in Fig.~\ref{fig:hist}.

\begin{figure}
	\centering
	\includegraphics[width=\columnwidth]{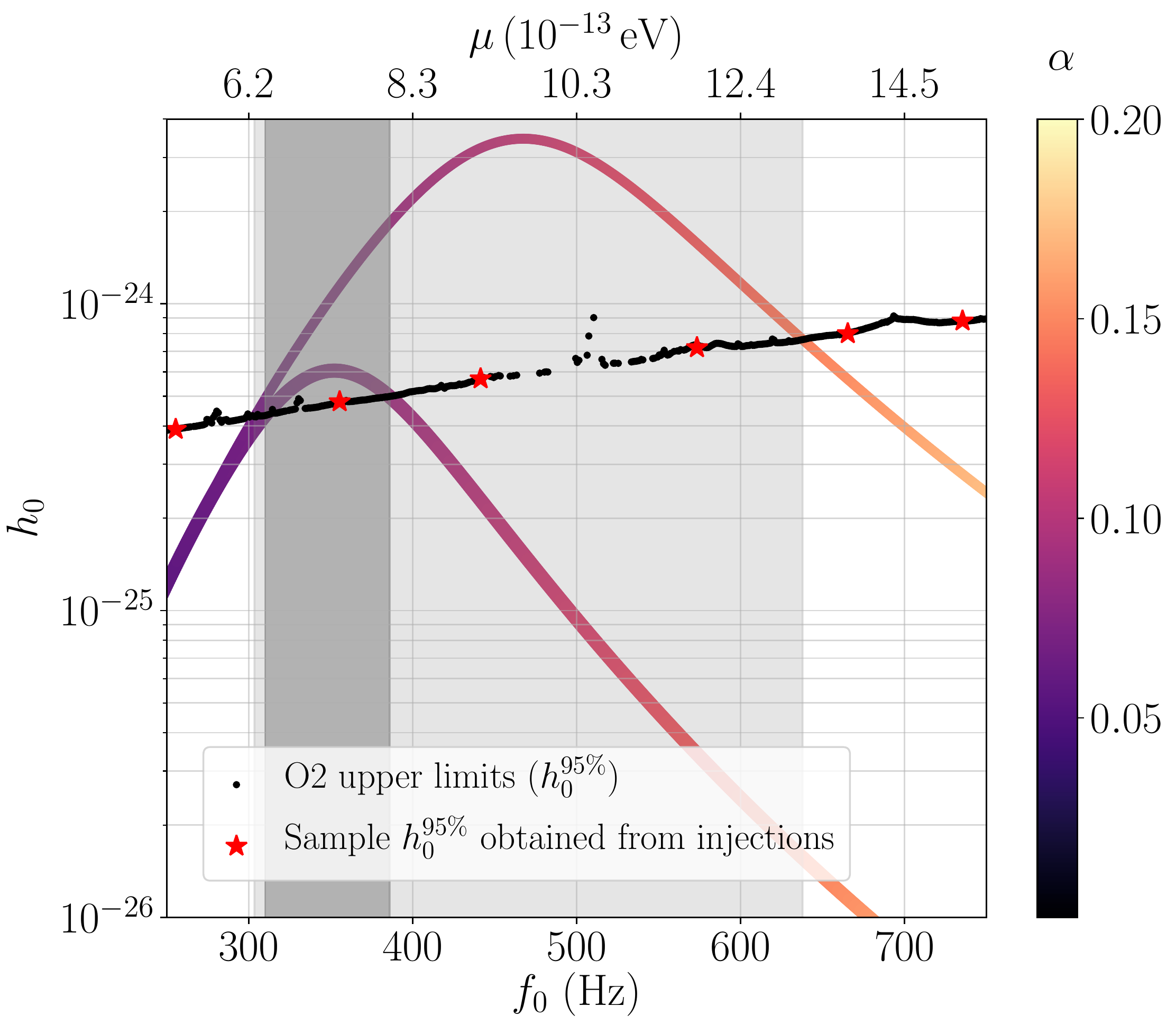}
	\caption[]{Frequentist strain upper limits at 95\% confidence ($h_0^{95\%}$) and disfavored scalar boson mass range. The colored curves show the numerically estimated signal strain ($h_0$) as a function of boson mass (top axis) and GW frequency (bottom axis). The thick and thin curves correspond to $t_{\rm age} = 5 \times 10^6$\,yr and $1 \times 10^5$\,yr, respectively. The color stands for the fine-structure constant ($\alpha$). The black dots indicate $h_0^{95\%}$ obtained from the search, assuming the electromagnetically measured orientation \mbox{$\iota = 27.1^\circ\pm0.8^\circ$}. The red stars mark $h_0^{95\%}$ obtained through injections in O2 data in six sample 1-Hz sub-bands. Sub-bands without a marker were vetoed. The shaded region marks the parameter space where $h_0^{95\%}$ beats the analytically estimated strain, and hence corresponds to the disfavored boson mass range without a detection: \mbox{$6.4 \leq \mu/(10^{-13}\,{\rm eV}) \leq 8.0$} for $t_{\rm age} = 5 \times 10^6$\,yr and \mbox{$6.3 \leq \mu/(10^{-13}\,{\rm eV}) \leq 13.2$} for $t_{\rm age} = 1 \times 10^5$\,yr. The source parameters adopted in the analytic estimation are $M = 14.8M_\odot$, $\chi_i=0.99$, and $d = 1.86$\,kpc.}
	\label{fig:h0_estimate}
\end{figure}

Unable to claim a detection, we adopt an empirical approach
to set a frequentist upper limit on $h_0$ at 95\% confidence ($h_0^{95\%}$).
Each black dot in Fig.~\ref{fig:h0_estimate} marks $h_0^{95\%}$ in the corresponding 1-Hz sub-band, derived from the O2 search assuming a source inclination $\iota = 27.1^\circ\pm0.8^\circ$ \cite{Orosz2011,Casares2014}.
The procedure for calculating $h_0^{95\%}$ is as follows. 
First, we perform Monte-Carlo simulations by injecting signals with a randomly chosen $f_0$ within 255--256\,Hz, but with a fixed $h_0$. We draw $\iota$ and $a_0$ uniformly within the ranges 26.3--27.9\,deg and 22.45--28.71\,l-s, respectively. We repeat this procedure for different $h_0$'s (step size $1 \times 10^{-26}$) until we find the value that yields a 95\% detection rate, viz.~$h_0^{95\%} = 3.9\times 10^{-25}$.
Next, we calculate $h_0^{95\%}$ over the full frequency band (black dots in Fig.~\ref{fig:h0_estimate}) using the analytical scaling $h_0^{95\%}(f) \propto S_h^{1/2}(f) f^{1/4}$ \cite{ScoX1ViterbiO1,Suvorova2016} \footnote{Summing the orbital sideband powers incoherently leads to the sensitivity loss scaling as $f^{1/4}$ \cite{Suvorova2016}.}, where $S_h(f)$ is the aLIGO O2 noise power spectral density \footnote{Here the effective $S_h(f)$ is calculated from the harmonic mean of the two detectors over all the 30-min short Fourier transforms collected from GPS time 1180310418 to 1187733592.}. 
At last, we verify the analytical scaling by repeating the first step in five other 1-Hz bands beginning at 355\,Hz, 441\,Hz, 573\,Hz, 665\,Hz, and 735\,Hz. The resulting $h_0^{95\%}$ values are marked by red stars in Fig.~\ref{fig:h0_estimate}.
The analytical scaling agrees with the empirical results in the sample sub-bands.

The statistical uncertainty of $h_0^{95\%}$ is less than ${\sim}2\%$, given that the step size of the injected signal strain amplitude is set to $1 \times 10^{-26}$.
Sub-bands containing a vetoed candidate are contaminated by instrumental artifacts. Hence we cannot place reliable upper limits in these bands (no black dot).

% \vspace{\baselineskip}
\subheading{Disfavored boson mass}
We assume that the scalar cloud is dominated by the energy level $l=m=1, n=0$, where $l$ and $n$ are the orbital azimuthal quantum number and radial quantum number, respectively, and the dominant GW mode is $l=m=2$.
In Fig.~\ref{fig:h0_estimate}, we plot the expected signal amplitude $h_0$ (colored curve) \cite{Isi2019} together with the upper-limit $h_0^{95\%}$ (black dots), as a function of expected signal frequency $f_0$ (bottom axis) and the corresponding boson mass (top axis).
To estimate $h_0$, we assume $M = 14.8M_\odot$, a distance $d = 1.86$\,kpc, and an age $t_{\rm age} = 5 \times 10^6$\,yr, based on current estimates for Cyg X-1 \cite{Orosz2011,Gou2011,Casares2014}. 
We also assume that the BH had an initial spin $\chi_i=0.99$ before the superradiant cloud growth.
The color bar indicates the system's ``gravitational fine structure constant'', defined as $\alpha = GM \omega_\mu/c^3$.

The shaded region in Fig.~\ref{fig:h0_estimate}, where the upper limits beat the estimated $h_0$, highlights the disfavored boson mass range.
The dark shaded region marks a conservatively disfavored mass range, \mbox{$6.4 \leq \mu/(10^{-13}\,{\rm eV}) \leq 8.0$}, corresponding to the conservative choice of the BH age~\footnote{The authors thank Ilya Mandel for helpful input regarding the potentially younger age of the BH.}, $t_{\rm age} = 5 \times 10^6$\,yr.
The estimated $h_0$ (thick colored curve) drops significantly for $\mu \gtrsim 7\times10^{-13}$\,eV ($\alpha \gtrsim 0.08$), because the timescale of the GW signal ($\tau_{\rm GW}$) depends strongly on $\alpha$ (Eqn.~(23) in Ref.~\cite{Isi2019}) and $h_0$ scales as $(1+t_{\rm age}/\tau_{\rm GW})^{-1}$ \cite{Arvanitaki2015}.
The cloud around an old BH with $t_{\rm age} \sim 5\times 10^6$\,yr, if it ever existed, would have mostly dissipated for $\alpha \gtrsim 0.1$.
If it is assumed that the BH is not much older than the X-ray binary jet, e.g., $t_{\rm age} = 10^5$\,yr~\cite{Russell2007}, the estimated $h_0$ is significantly larger for $\mu \gtrsim 6.2\times 10^{-13}$~eV (thin colored curve), and a wider mass range of \mbox{$6.3 \leq \mu/(10^{-13}\,{\rm eV}) \leq 13.2$} is disfavored (light shaded region) \cite{Sun2020-erratum}.
In Fig.~\ref{fig:h0_estimate}, we do not rely on Cyg X-1 spin measurements and rather let the post-superradiance spin of the BH ($\chi_f$) be a free parameter.
In the frequency band searched, the conjectured cloud would have spun down the BH such that $\chi_f \lesssim 0.6$ ($0.3 \lesssim \chi_f \lesssim 0.4$ in the shaded region).
We do not expect the accreting matter to torque up the BH, since the spin-up timescale is on the order of $10^7$\,yrs ($>t_{\rm age}$) even at the Eddington accretion rate \cite{Gou2011}, much longer than the timescale of the superradiant instability ($\sim$\,yrs).
If we refer to existing measurements for Cyg X-1 and assume $\chi_f \geq 0.95$ \cite{Gou2011}, we obtain $\tau_{\rm GW} \ll t_{\rm age}$ and $f_0 > 1.5$\,kHz.
The existing method cannot handle the widely spread sidebands ($>1$\,Hz) at such high frequency.
Even if the search could be extended to $f_0 > 1.5$\,kHz with an improved method, the corresponding boson mass ($\mu > 3\times 10^{-12}$\,eV) cannot be excluded in the absence of a detection, since a signal with $\tau_{\rm GW} \ll t_{\rm age}$ would no longer be present.

The uncertainty in our constraints is dominated by uncertainties in the source properties. The measurement uncertainties of the BH mass, initial spin, and age could lead to an error of at most ${\sim}30\%$ on the expected $h_0$. The statistical uncertainties of the expected $h_0$ from numerical modeling and the upper-limit $h_0^{95\%}$ are both on the order of a couple of percent.

% \vspace{\baselineskip}
\subheading{String axiverse}
In the discussion above, we implicitly assume that the boson does not self-interact significantly.
\citet{Yoshino2015} studied the superradiant instability in the string axiverse scenario, taking into consideration the nonlinear self-interaction of string axions.
As the scalar field $\Phi$ grows through the superradiant instability and reaches a level of $\Phi \sim f_a$, the nonlinear self-interaction triggers a ``bosenova", i.e., the axion cloud partially collapses, with about 5\% energy falling back into the BH \cite{Yoshino2012,Yoshino:2015nsa,Yoshino2015}.
After abruptly dropping, the field restarts its superradiant growth until the ``bosenova'' is triggered again.
For bosons that can be probed with aLIGO and Virgo, this process may occur for $f_a$ values smaller than the GUT scale.
It was suggested that this periodic process could prevent superradiance from being saturated, allowing the presence of a string-axion cloud around an old, high-spin BH like Cyg X-1.

\begin{figure}
	\centering
	\includegraphics[width=\columnwidth]{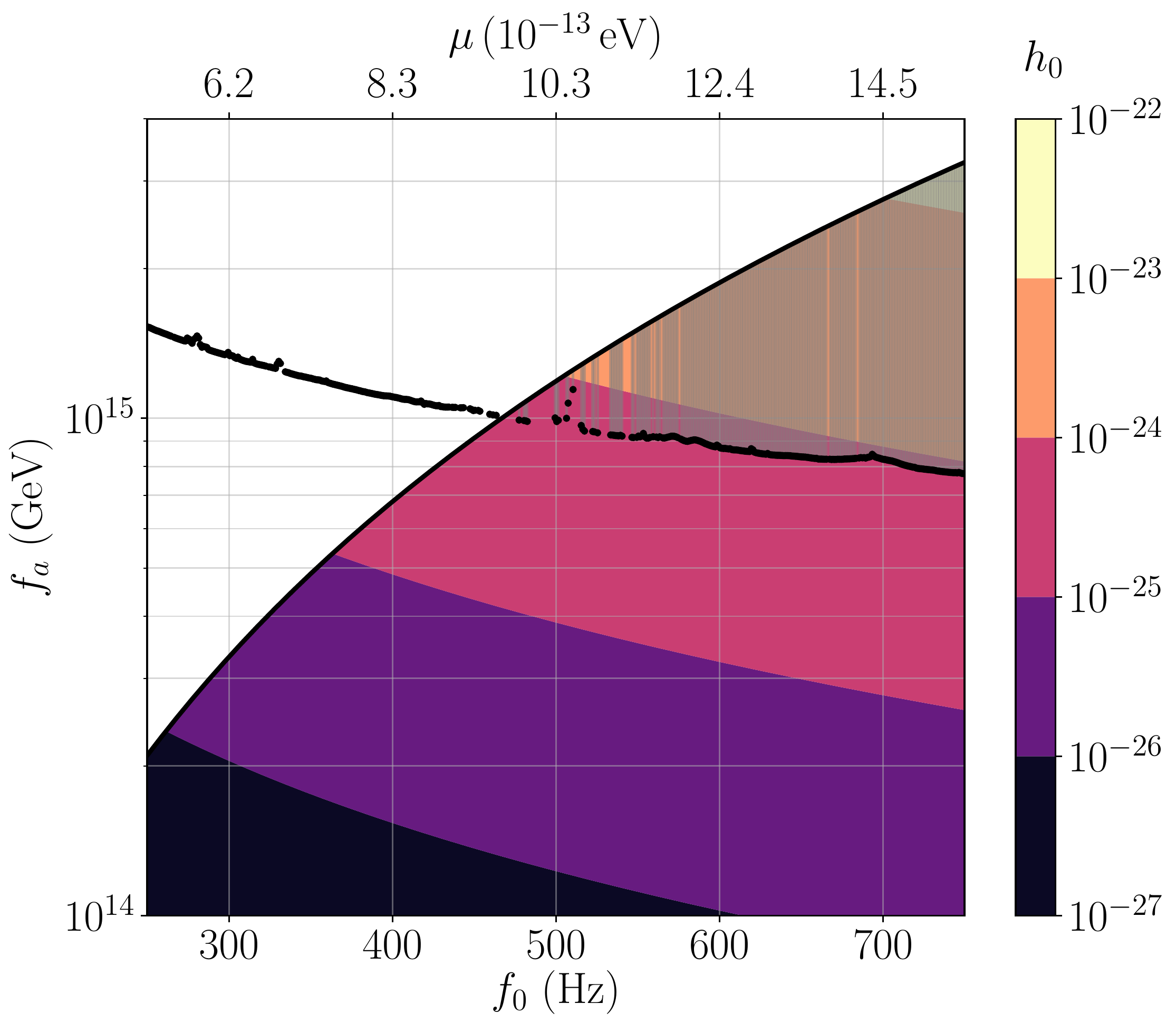}
	\caption[]{Excluded parameter space of the decay constant $f_a$ and mass $\mu$ of string axions. The black dots indicate the upper limits on $f_a$ as a function of $\mu$ (top axis) and $f_0$ (bottom axis) obtained from the search. The contours indicate the estimated $h_0$. The white region is the parameter space where the condition for the bosenova to occur is not satisfied. The shaded region indicates the excluded parameter space.}
	\label{fig:h0_axiverse}
\end{figure}

We may constrain the above scenario by comparing our upper limits $h_0^{95\%}$ to the estimated strain for a cloud that saturates the threshold for the bosenova to occur, namely \cite{Yoshino2015}
\begin{eqnarray}
	\nonumber
	h_0 & \approx & 6.2 \times 10^{-25} \left(\frac{f_a}{10^{16}\,{\rm GeV}}\right)^2 \left(\frac{\mu}{10^{-13}\,{\rm eV}}\right)^2 \\
	&&\times \left(\frac{M}{14.8M_\odot}\right)^3 \left( \frac{1.86\,{\rm kpc}}{d} \right) .
\end{eqnarray}
The results are displayed in the ($f_a, \mu$) plane in Fig.~\ref{fig:h0_axiverse}. The black dots are calculated from $h_0^{95\%}$, and the contours indicate the estimated $h_0$ values. The boundary between the white and colored regions corresponds to
\begin{equation}
\left(\frac{f_a}{10^{15}\,{\rm GeV}}\right) = 3.5\times 10^{-3}\left(\frac{\mu}{10^{-13}\,{\rm eV}}\right)^{5/2} \left(\frac{M}{14.8M_\odot}\right)^{5/2}, 
\end{equation}
below which a scalar field grown out of the superradiant instability can reach the level $\Phi/f_a \gtrsim 0.67$ to trigger the bosenova \cite{Yoshino2012,Yoshino:2015nsa,Yoshino2015}.
Here, we have assumed \mbox{$\chi_i = 0.99$} and \mbox{$\chi_f = 0.95$}, consistent with the Cyg X-1 observations.
We do not derive constraints in the white region, where the condition for the bosenova to occur is not satisfied.
The shaded region highlights the ($f_a, \mu$) space excluded by the search.
In the mass range \mbox{$9.6 \leq \mu/(10^{-13}\,{\rm eV}) \leq 15.5$}, the measurement excludes $f_a\sim 1\times 10^{15}$\,GeV, an order of magnitude smaller than the GUT scale.

These constraints are contingent on the signal lifetime being longer than $t_{\rm age}$ in this model.
Furthermore, this scenario does not take into account the potential impact of radial cloud oscillations: the cloud expands and shrinks during the superradiance and bosenova processes, respectively, potentially modulating the GW frequency on a timescale of minutes \cite{Yoshino2015}.
Accurate estimates of signal duration and the frequency modulation will require further numerical study.

% \vspace{\baselineskip}
\heading{Conclusion}
We report the results from a directed search for GW signals from a putative scalar boson cloud around the BH in Cyg X-1 in the aLIGO O2 run, using an efficient HMM tracking scheme and a frequency domain matched filter.
%The Doppler modulation of the signal caused by the binary motion is accounted for.
We find no evidence of GW signals in the frequency band 250--750\,Hz. 
Assuming an age of $5 \times 10^6$\,yr, a nearly-extremal spin at birth, and an unknown post-superradiance BH spin, our measurement disfavors scalar boson masses in \mbox{$6.4 \leq \mu/(10^{-13}\,{\rm eV}) \leq 8.0$}.
No reliable constraint can be placed for $\chi_f \geq 0.95$ without considering particle self-interactions, since the boson field would have disappeared in less than ${\sim}10^6$\,yr.
On the other hand, in the string axiverse scenario, the axion's self-interactions could prevent superradiance from being saturated, enabling the existence of a cloud around an old BH with high spin.
We can thus exclude $f_a \sim 1\times 10^{15}$\,GeV for string axions in the mass range \mbox{$9.6 \leq \mu/(10^{-13}\,{\rm eV}) \leq 15.5$}.
This assumes that $\tau_{\rm GW} > t_{\rm age}$, and that the frequency modulation from the cloud oscillations does not impact the search sensitivity.
A more robust analysis will be possible when numerical results of the GW signal timescale and waveform become available under this model. 
In both scenarios, constraints can only be derived for frequencies where the estimated signal strain exceeds the upper limit obtained from the search. Analyzing a broader frequency band would not have improved the obtained boson mass constraints. 
 
This is a first dedicated GW search for ultralight bosons targeting a known BH. It demonstrates the methodology and interpretation for future searches of similar kind.
Besides X-ray binaries, when nearby, well localized CBCs are detected in upcoming observing runs, the young, isolated remnant BHs will be a target of great interest, free of the complications related to BH age and orbital motion.
Future detectors promise to enable further boson constraints, or even a detection \cite{Isi2019,Ghosh:2018gaw,Hild:2010id,Sathyaprakash:2012jk,Punturo:2010zz,Abbott2017-nextGen-CE}. 

% \vspace{\baselineskip}
%\heading{Acknowledgments}
\begin{acknowledgments}
We are grateful to Marianne Heida and Riley M. Connors for discussions and suggestions about the source parameters, and the LIGO and Virgo Continuous Wave Working Group for discussions and comments.
% LIGO
This search uses LIGO data from the Gravitational Wave Open Science Center (https://www.gw-openscience.org).
The authors are grateful for computational resources provided by the LIGO Laboratory.
LIGO was constructed by the California Institute of Technology and Massachusetts Institute of Technology with funding from the National Science Foundation, and operates under cooperative agreement PHY--0757058. Advanced LIGO was built under award PHY--0823459. 
% NASA
Support for this work was provided by NASA through the NASA Hubble Fellowship grant No. HST--HF2--51410.001--A awarded by the Space Telescope Science Institute, which is operated by the Association of Universities for Research in Astronomy, Inc., for NASA, under contract NAS5--26555.
% Richard
R. Brito acknowledges financial support from the European Union's Horizon 2020 research and innovation programme under the Marie Sk\l odowska-Curie grant agreement No. 792862.
% Aspen
This work was written in part at the Aspen Center for Physics, which is supported by National Science Foundation grant PHY--1607611.
This paper carries LIGO Document No.~\dcc.
\end{acknowledgments}

% \twocolumngrid  \vspace{1cm} 
% \begin{center}  
% 	{\large\bf Supplemental Material} 
% \end{center} 
\appendix

\section{Hidden Markov model}
\label{sec:hmm}
A HMM is a finite state automaton defined by a hidden (unobservable) state variable $q(t_k) \in \{q_1, \cdots, q_{N_Q}\}$ and an observable state variable $o(t_k) \in \{o_1, \cdots, o_{N_O}\}$ at discrete times $t_k \in \{t_0, \cdots, t_{N_T}\}$. The automaton is observed in the state $o_j$ with emission probability \cite{Suvorova2016}
\begin{equation}
\label{eqn:likelihood}
L_{o_j q_i} = \Pr [o(t_k)=o_j|q(t_k)=q_i],
\end{equation}
and jumps between hidden states from $t_k$ to $t_{k+1}$ with transition probability \cite{Suvorova2016}
\begin{equation}
\label{eqn:prob_matrix}
A_{q_j q_i} = \Pr [q(t_{k+1})=q_j|q(t_k)=q_i].
\end{equation}

For a memoryless Markov process, the probability that the hidden path $Q=\{q(t_0), \cdots, q(t_{N_T})\}$ gives rise to the observed sequence $O=\{o(t_0), \cdots, o(t_{N_T})\}$ is given by \cite{Suvorova2016}
\begin{equation}
	\label{eqn:prob}
	\begin{split}
		\Pr(Q|O) = & L_{o(t_{N_T})q(t_{N_T})} A_{q(t_{N_T})q(t_{N_T-1})} \cdots L_{o(t_1)q(t_1)} \\ 
		& \times A_{q(t_1)q(t_0)} \Pi_{q(t_0)},
	\end{split}
\end{equation}
where 
\begin{equation}
	\Pi_{q_i} = \Pr [q(t_0)=q_i]
\end{equation}
is the prior.
The most probable path $Q^*(O)$, maximizing $\Pr(Q|O)$, gives the best estimate of $q(t_k)$ over the total observing time.

Discrete hidden states $q_i$ ($1\leq i \leq N_Q$) are mapped one-to-one to the frequency bins in the output of a frequency-domain estimator $G(f)$ computed over $\Tcoh$, with bin size $\Delta f = 1/(2 \Tcoh)$. We choose $\Tcoh$ to satisfy
\begin{equation}
	\left|\int_t^{t+\Tcoh}dt' \dot{f_0}(t')\right| < \Delta f,
\end{equation}
for $0\leq t \leq (\Tobs-\Tcoh)$, such that searching over $\dot{f_0}$ is not necessary. For the signal model considered in Ref.~\cite{Isi2019}, Eqn.~(\ref{eqn:likelihood}) can be written as
\begin{equation}
\label{eqn:emission_prob}
L_{o(t_k) f_i} \propto \exp[G(f_i)],
\end{equation}
where $G(f_i)$ is the log likelihood that the signal frequency $f_0$ lies in bin $[f_i, f_i +\Delta f]$ during interval $[t_k, t_k+\Tcoh]$. Eqn.~(\ref{eqn:prob_matrix}) takes the simplified form
\begin{equation}
	\label{eqn:trans_matrix}
	A_{f_{i+1} f_i} = A_{f_i f_i} = \frac{1}{2},
\end{equation}
with all other entries vanishing. The prior is set to \mbox{$\Pi_{f_i} = N_Q^{-1}$} as we have no advance knowledge of $f_0$.

\section{Viterbi algorithm and detection score}
\label{sec:viterbi}
The Viterbi algorithm is used to compute $Q^*(O)$ recursively \cite{Viterbi1967, Quinn2001}. At every step $k$ ($1\leq k \leq N_T$) forward, the algorithm only keeps $N_Q$ possible state sequences ending in state $q_i$ ($1\leq i \leq N_Q$), and stores their maximum probabilities \cite{Suvorova2016}
\begin{equation}
	\delta_{q_i}(t_k) = L_{o(t_k)q_i} \mathop{\max} \limits_{1 \leq j \leq N_Q} [A_{q_i q_j}\delta_{q_j}(t_{k-1})],
\end{equation}
as well as the previous-step states of origin \cite{Suvorova2016},
\begin{equation}
	\Phi_{q_i}(t_k) = \mathop{\arg \max} \limits_{1 \leq j \leq N_Q} [A_{q_i q_j}\delta_{q_j}(t_{k-1})],
\end{equation}
that maximize the probabilities at step $k$. The optimal Viterbi path is then reconstructed by backtracking
\begin{equation}
	q^*(t_k) = \Phi_{q^*(t_{k+1})}(t_{k+1}),
\end{equation}
for $N_T -1 \geq k \geq 0 $. 

In this application, the detection score $S$ is defined, such that the log likelihood of the optimal Viterbi path equals the mean log likelihood of all paths plus $S$ standard deviations, viz. \cite{ScoX1ViterbiO1}
\begin{equation}
	S = \frac{\ln \delta_{f_0^*}{(t_{N_T})} -\mu_{\ln \delta}(t_{N_T})}{\sigma_{\ln \delta}(t_{N_T})},
\end{equation}
with
\begin{equation}
	\mu_{\ln \delta}(t_{N_T}) = N_Q^{-1} \sum_{i=1}^{N_Q} \ln \delta_{f_i}(t_{N_T}),
\end{equation}
and
\begin{equation}
	\sigma_{\ln \delta}(t_{N_T})^2 = N_Q^{-1} \sum_{i=1}^{N_Q} [\ln \delta_{f_i}(t_{N_T}) - \mu_{\ln \delta}(t_{N_T}) ]^2,
\end{equation}
where $\delta_{f_i}(t_{N_T})$ denotes the maximum probability of the frequency path ending in bin $i$ ($1\leq i \leq N_Q$) at step $N_T$, and $\delta_{f_0^*}{(t_{N_T})}$ is the likelihood of the optimal Viterbi path.

\section{Matched filter: Bessel-weighted $\mathcal{F}$-statistic}
\label{sec:matched_filter}
The optimal frequency-domain matched filter for a continuous-wave signal with no orbital motion is the maximum-likelihood $\mathcal{F}$-statistic, $\mathcal{F}(f)$ \cite{Jaranowski1998,F-stat2019}, which accounts for the Earth's motion with respect to the Solar System barycenter (SSB). When the source orbits a companion, the GW signal frequency is modulated due to the orbital Doppler effect. For a Keplerian circular orbit, the GW strain can be expanded in a Jacobi-Anger series as \cite{Abramowitz1964,Suvorova2016}
\begin{equation}
\label{eqn:wave_strain_expansion}
h(t) \propto \mathop{\sum} \limits_{n=-\infty}^{\infty} J_n(2 \pi f_0 a_0) \cos [2\pi (f_0 + n/P)t],
\end{equation}
where $J_n(z)$ is a Bessel function of order $n$ of the first kind.
The $\mathcal{F}$-statistic power is distributed into approximately $M=2\,\text{ceil} (2 \pi f_0 a_0)+1$ orbital sidebands, separated by $1/P$, where $\text{ceil}(x)$ denotes the smallest integer greater than or equal to $x$.
Hence we use $G(f)=\mathcal{F}(f) \otimes B(f)$, a Bessel-weighted $\mathcal{F}$-statistic, in Eqn.~(\ref{eqn:emission_prob}) for a source in a binary orbit, where $B(f)$ is given by \cite{Suvorova2016}
\begin{equation}
B(f)=\sum\limits_{n=-(M-1)/2}^{(M-1)/2} [J_n(2 \pi f a_0)]^2 \delta(f-n/P).
\end{equation}


\begin{thebibliography}{68}%
	\makeatletter
	\providecommand \@ifxundefined [1]{%
		\@ifx{#1\undefined}
	}%
	\providecommand \@ifnum [1]{%
		\ifnum #1\expandafter \@firstoftwo
		\else \expandafter \@secondoftwo
		\fi
	}%
	\providecommand \@ifx [1]{%
		\ifx #1\expandafter \@firstoftwo
		\else \expandafter \@secondoftwo
		\fi
	}%
	\providecommand \natexlab [1]{#1}%
	\providecommand \enquote  [1]{``#1''}%
	\providecommand \bibnamefont  [1]{#1}%
	\providecommand \bibfnamefont [1]{#1}%
	\providecommand \citenamefont [1]{#1}%
	\providecommand \href@noop [0]{\@secondoftwo}%
	\providecommand \href [0]{\begingroup \@sanitize@url \@href}%
	\providecommand \@href[1]{\@@startlink{#1}\@@href}%
	\providecommand \@@href[1]{\endgroup#1\@@endlink}%
	\providecommand \@sanitize@url [0]{\catcode `\\12\catcode `\$12\catcode
		`\&12\catcode `\#12\catcode `\^12\catcode `\_12\catcode `\%12\relax}%
	\providecommand \@@startlink[1]{}%
	\providecommand \@@endlink[0]{}%
	\providecommand \url  [0]{\begingroup\@sanitize@url \@url }%
	\providecommand \@url [1]{\endgroup\@href {#1}{\urlprefix }}%
	\providecommand \urlprefix  [0]{URL }%
	\providecommand \Eprint [0]{\href }%
	\providecommand \doibase [0]{http://dx.doi.org/}%
	\providecommand \selectlanguage [0]{\@gobble}%
	\providecommand \bibinfo  [0]{\@secondoftwo}%
	\providecommand \bibfield  [0]{\@secondoftwo}%
	\providecommand \translation [1]{[#1]}%
	\providecommand \BibitemOpen [0]{}%
	\providecommand \bibitemStop [0]{}%
	\providecommand \bibitemNoStop [0]{.\EOS\space}%
	\providecommand \EOS [0]{\spacefactor3000\relax}%
	\providecommand \BibitemShut  [1]{\csname bibitem#1\endcsname}%
	\let\auto@bib@innerbib\@empty
	%</preamble>
	\bibitem [{\citenamefont {{Peccei}}\ and\ \citenamefont
		{{Quinn}}(1977{\natexlab{a}})}]{Peccei1977}%
	\BibitemOpen
	\bibfield  {author} {\bibinfo {author} {\bibfnamefont {R.~D.}\ \bibnamefont
			{{Peccei}}}\ and\ \bibinfo {author} {\bibfnamefont {H.~R.}\ \bibnamefont
			{{Quinn}}},\ }\bibfield  {title} {\enquote {\bibinfo {title} {{CP
					conservation in the presence of pseudoparticles}},}\ }\href {\doibase
		10.1103/PhysRevLett.38.1440} {\bibfield  {journal} {\bibinfo  {journal}
			{Physical Review Letters}\ }\textbf {\bibinfo {volume} {38}},\ \bibinfo
		{pages} {1440--1443} (\bibinfo {year} {1977}{\natexlab{a}})}\BibitemShut
	{NoStop}%
	\bibitem [{\citenamefont {{Peccei}}\ and\ \citenamefont
		{{Quinn}}(1977{\natexlab{b}})}]{Peccei1977PhRvD}%
	\BibitemOpen
	\bibfield  {author} {\bibinfo {author} {\bibfnamefont {R.~D.}\ \bibnamefont
			{{Peccei}}}\ and\ \bibinfo {author} {\bibfnamefont {H.~R.}\ \bibnamefont
			{{Quinn}}},\ }\bibfield  {title} {\enquote {\bibinfo {title} {{Constraints
					imposed by CP conservation in the presence of pseudoparticles}},}\ }\href
	{\doibase 10.1103/PhysRevD.16.1791} {\bibfield  {journal} {\bibinfo
			{journal} {Physical Review D}\ }\textbf {\bibinfo {volume} {16}},\ \bibinfo
		{pages} {1791--1797} (\bibinfo {year} {1977}{\natexlab{b}})}\BibitemShut
	{NoStop}%
	\bibitem [{\citenamefont {{Weinberg}}(1978)}]{Weinberg1978}%
	\BibitemOpen
	\bibfield  {author} {\bibinfo {author} {\bibfnamefont {Steven}\ \bibnamefont
			{{Weinberg}}},\ }\bibfield  {title} {\enquote {\bibinfo {title} {{A new light
					boson?}}}\ }\href {\doibase 10.1103/PhysRevLett.40.223} {\bibfield  {journal}
		{\bibinfo  {journal} {Physical Review Letters}\ }\textbf {\bibinfo {volume}
			{40}},\ \bibinfo {pages} {223--226} (\bibinfo {year} {1978})}\BibitemShut
	{NoStop}%
	\bibitem [{\citenamefont {Arvanitaki}\ \emph {et~al.}(2010)\citenamefont
		{Arvanitaki}, \citenamefont {Dimopoulos}, \citenamefont {Dubovsky},
		\citenamefont {Kaloper},\ and\ \citenamefont
		{March-Russell}}]{Arvanitaki2010}%
	\BibitemOpen
	\bibfield  {author} {\bibinfo {author} {\bibfnamefont {Asimina}\ \bibnamefont
			{Arvanitaki}}, \bibinfo {author} {\bibfnamefont {Savas}\ \bibnamefont
			{Dimopoulos}}, \bibinfo {author} {\bibfnamefont {Sergei}\ \bibnamefont
			{Dubovsky}}, \bibinfo {author} {\bibfnamefont {Nemanja}\ \bibnamefont
			{Kaloper}}, \ and\ \bibinfo {author} {\bibfnamefont {John}\ \bibnamefont
			{March-Russell}},\ }\bibfield  {title} {\enquote {\bibinfo {title} {{String
					axiverse}},}\ }\href {\doibase 10.1103/PhysRevD.81.123530} {\bibfield
		{journal} {\bibinfo  {journal} {Physical Review D}\ }\textbf {\bibinfo
			{volume} {81}},\ \bibinfo {pages} {123530} (\bibinfo {year}
		{2010})}\BibitemShut {NoStop}%
	\bibitem [{\citenamefont {Goodsell}\ \emph {et~al.}(2009)\citenamefont
		{Goodsell}, \citenamefont {Jaeckel}, \citenamefont {Redondo},\ and\
		\citenamefont {Ringwald}}]{Goodsell2009}%
	\BibitemOpen
	\bibfield  {author} {\bibinfo {author} {\bibfnamefont {Mark}\ \bibnamefont
			{Goodsell}}, \bibinfo {author} {\bibfnamefont {Joerg}\ \bibnamefont
			{Jaeckel}}, \bibinfo {author} {\bibfnamefont {Javier}\ \bibnamefont
			{Redondo}}, \ and\ \bibinfo {author} {\bibfnamefont {Andreas}\ \bibnamefont
			{Ringwald}},\ }\bibfield  {title} {\enquote {\bibinfo {title} {{Naturally
					Light Hidden Photons in LARGE Volume String Compactifications}},}\ }\href
	{\doibase 10.1088/1126-6708/2009/11/027} {\bibfield  {journal} {\bibinfo
			{journal} {JHEP}\ }\textbf {\bibinfo {volume} {11}},\ \bibinfo {pages} {027}
		(\bibinfo {year} {2009})}\BibitemShut {NoStop}%
	\bibitem [{\citenamefont {Jaeckel}\ and\ \citenamefont
		{Ringwald}(2010)}]{Jaeckel:2010ni}%
	\BibitemOpen
	\bibfield  {author} {\bibinfo {author} {\bibfnamefont {Joerg}\ \bibnamefont
			{Jaeckel}}\ and\ \bibinfo {author} {\bibfnamefont {Andreas}\ \bibnamefont
			{Ringwald}},\ }\bibfield  {title} {\enquote {\bibinfo {title} {{The
					Low-Energy Frontier of Particle Physics}},}\ }\href {\doibase
		10.1146/annurev.nucl.012809.104433} {\bibfield  {journal} {\bibinfo
			{journal} {Ann. Rev. Nucl. Part. Sci.}\ }\textbf {\bibinfo {volume} {60}},\
		\bibinfo {pages} {405--437} (\bibinfo {year} {2010})}\BibitemShut {NoStop}%
	\bibitem [{\citenamefont {Essig}\ \emph {et~al.}(2013)\citenamefont {Essig}
		\emph {et~al.}}]{Essig:2013lka}%
	\BibitemOpen
	\bibfield  {author} {\bibinfo {author} {\bibfnamefont {Rouven}\ \bibnamefont
			{Essig}} \emph {et~al.},\ }\bibfield  {title} {\enquote {\bibinfo {title}
			{{Working Group Report: New Light Weakly Coupled Particles}},}\ }in\ \href
	{http://www.slac.stanford.edu/econf/C1307292/docs/
		IntensityFrontier/NewLight-17.pdf} {\emph {\bibinfo {booktitle}
			{{Proceedings, 2013 Community Summer Study on the Future of U.S. Particle
					Physics: Snowmass on the Mississippi (CSS2013): Minneapolis, MN, USA, July
					29-August 6, 2013}}}}\ (\bibinfo {year} {2013})\BibitemShut {NoStop}%
	\bibitem [{\citenamefont {Hui}\ \emph {et~al.}(2017)\citenamefont {Hui},
		\citenamefont {Ostriker}, \citenamefont {Tremaine},\ and\ \citenamefont
		{Witten}}]{Hui:2016ltb}%
	\BibitemOpen
	\bibfield  {author} {\bibinfo {author} {\bibfnamefont {Lam}\ \bibnamefont
			{Hui}}, \bibinfo {author} {\bibfnamefont {Jeremiah~P.}\ \bibnamefont
			{Ostriker}}, \bibinfo {author} {\bibfnamefont {Scott}\ \bibnamefont
			{Tremaine}}, \ and\ \bibinfo {author} {\bibfnamefont {Edward}\ \bibnamefont
			{Witten}},\ }\bibfield  {title} {\enquote {\bibinfo {title} {{Ultralight
					scalars as cosmological dark matter}},}\ }\href {\doibase
		10.1103/PhysRevD.95.043541} {\bibfield  {journal} {\bibinfo  {journal} {Phys.
				Rev.}\ }\textbf {\bibinfo {volume} {D95}},\ \bibinfo {pages} {043541}
		(\bibinfo {year} {2017})}\BibitemShut {NoStop}%
	\bibitem [{\citenamefont {Arvanitaki}\ and\ \citenamefont
		{Dubovsky}(2011)}]{Arvanitaki2011}%
	\BibitemOpen
	\bibfield  {author} {\bibinfo {author} {\bibfnamefont {Asimina}\ \bibnamefont
			{Arvanitaki}}\ and\ \bibinfo {author} {\bibfnamefont {Sergei}\ \bibnamefont
			{Dubovsky}},\ }\bibfield  {title} {\enquote {\bibinfo {title} {{Exploring the
					string axiverse with precision black hole physics}},}\ }\href {\doibase
		10.1103/PhysRevD.83.044026} {\bibfield  {journal} {\bibinfo  {journal}
			{Physical Review D}\ }\textbf {\bibinfo {volume} {83}},\ \bibinfo {pages}
		{044026} (\bibinfo {year} {2011})}\BibitemShut {NoStop}%
	\bibitem [{\citenamefont {Yoshino}\ and\ \citenamefont
		{Kodama}(2014)}]{Yoshino2014}%
	\BibitemOpen
	\bibfield  {author} {\bibinfo {author} {\bibfnamefont {Hirotaka}\
			\bibnamefont {Yoshino}}\ and\ \bibinfo {author} {\bibfnamefont {Hideo}\
			\bibnamefont {Kodama}},\ }\bibfield  {title} {\enquote {\bibinfo {title}
			{{Gravitational radiation from an axion cloud around a black hole:
					Superradiant phase}},}\ }\href {\doibase 10.1093/ptep/ptu029} {\bibfield
		{journal} {\bibinfo  {journal} {Progress of Theoretical and Experimental
				Physics}\ }\textbf {\bibinfo {volume} {2014}},\ \bibinfo {pages} {43E02--0}
		(\bibinfo {year} {2014})}\BibitemShut {NoStop}%
	\bibitem [{\citenamefont {Yoshino}\ and\ \citenamefont
		{Kodama}(2015{\natexlab{a}})}]{Yoshino2015}%
	\BibitemOpen
	\bibfield  {author} {\bibinfo {author} {\bibfnamefont {Hirotaka}\
			\bibnamefont {Yoshino}}\ and\ \bibinfo {author} {\bibfnamefont {Hideo}\
			\bibnamefont {Kodama}},\ }\bibfield  {title} {\enquote {\bibinfo {title}
			{{Probing the string axiverse by gravitational waves from Cygnus X-1}},}\
	}\href {\doibase 10.1093/ptep/ptv067} {\bibfield  {journal} {\bibinfo
			{journal} {Progress of Theoretical and Experimental Physics}\ }\textbf
		{\bibinfo {volume} {2015}} (\bibinfo {year} {2015}{\natexlab{a}}),\
		10.1093/ptep/ptv067}\BibitemShut {NoStop}%
	\bibitem [{\citenamefont {Arvanitaki}\ \emph {et~al.}(2015)\citenamefont
		{Arvanitaki}, \citenamefont {Baryakhtar},\ and\ \citenamefont
		{Huang}}]{Arvanitaki2015}%
	\BibitemOpen
	\bibfield  {author} {\bibinfo {author} {\bibfnamefont {Asimina}\ \bibnamefont
			{Arvanitaki}}, \bibinfo {author} {\bibfnamefont {Masha}\ \bibnamefont
			{Baryakhtar}}, \ and\ \bibinfo {author} {\bibfnamefont {Xinlu}\ \bibnamefont
			{Huang}},\ }\bibfield  {title} {\enquote {\bibinfo {title} {{Discovering the
					QCD axion with black holes and gravitational waves}},}\ }\href {\doibase
		10.1103/PhysRevD.91.084011} {\bibfield  {journal} {\bibinfo  {journal}
			{Physical Review D}\ }\textbf {\bibinfo {volume} {91}},\ \bibinfo {pages}
		{084011} (\bibinfo {year} {2015})}\BibitemShut {NoStop}%
	\bibitem [{\citenamefont {Arvanitaki}\ \emph {et~al.}(2017)\citenamefont
		{Arvanitaki}, \citenamefont {Baryakhtar}, \citenamefont {Dimopoulos},
		\citenamefont {Dubovsky},\ and\ \citenamefont {Lasenby}}]{Arvanitaki2017}%
	\BibitemOpen
	\bibfield  {author} {\bibinfo {author} {\bibfnamefont {Asimina}\ \bibnamefont
			{Arvanitaki}}, \bibinfo {author} {\bibfnamefont {Masha}\ \bibnamefont
			{Baryakhtar}}, \bibinfo {author} {\bibfnamefont {Savas}\ \bibnamefont
			{Dimopoulos}}, \bibinfo {author} {\bibfnamefont {Sergei}\ \bibnamefont
			{Dubovsky}}, \ and\ \bibinfo {author} {\bibfnamefont {Robert}\ \bibnamefont
			{Lasenby}},\ }\bibfield  {title} {\enquote {\bibinfo {title} {{Black hole
					mergers and the QCD axion at Advanced LIGO}},}\ }\href {\doibase
		10.1103/PhysRevD.95.043001} {\bibfield  {journal} {\bibinfo  {journal}
			{Physical Review D}\ }\textbf {\bibinfo {volume} {95}},\ \bibinfo {pages}
		{043001} (\bibinfo {year} {2017})}\BibitemShut {NoStop}%
	\bibitem [{\citenamefont {Brito}\ \emph
		{et~al.}(2017{\natexlab{a}})\citenamefont {Brito}, \citenamefont {{Ghosh}},
		\citenamefont {{Barausse}}, \citenamefont {{Berti}}, \citenamefont
		{{Cardoso}}, \citenamefont {{Dvorkin}}, \citenamefont {{Klein}},\ and\
		\citenamefont {{Pani}}}]{Brito2017-letter}%
	\BibitemOpen
	\bibfield  {author} {\bibinfo {author} {\bibfnamefont {Richard}\ \bibnamefont
			{Brito}}, \bibinfo {author} {\bibfnamefont {S.}~\bibnamefont {{Ghosh}}},
		\bibinfo {author} {\bibfnamefont {E.}~\bibnamefont {{Barausse}}}, \bibinfo
		{author} {\bibfnamefont {E.}~\bibnamefont {{Berti}}}, \bibinfo {author}
		{\bibfnamefont {V.}~\bibnamefont {{Cardoso}}}, \bibinfo {author}
		{\bibfnamefont {I.}~\bibnamefont {{Dvorkin}}}, \bibinfo {author}
		{\bibfnamefont {A.}~\bibnamefont {{Klein}}}, \ and\ \bibinfo {author}
		{\bibfnamefont {P.}~\bibnamefont {{Pani}}},\ }\bibfield  {title} {\enquote
		{\bibinfo {title} {{Stochastic and Resolvable Gravitational Waves from
					Ultralight Bosons}},}\ }\href {\doibase 10.1103/PhysRevLett.119.131101}
	{\bibfield  {journal} {\bibinfo  {journal} {Physical Review Letters}\
		}\textbf {\bibinfo {volume} {119}},\ \bibinfo {pages} {131101} (\bibinfo
		{year} {2017}{\natexlab{a}})}\BibitemShut {NoStop}%
	\bibitem [{\citenamefont {Brito}\ \emph
		{et~al.}(2017{\natexlab{b}})\citenamefont {Brito}, \citenamefont {{Ghosh}},
		\citenamefont {{Barausse}}, \citenamefont {{Berti}}, \citenamefont
		{{Cardoso}}, \citenamefont {{Dvorkin}}, \citenamefont {{Klein}},\ and\
		\citenamefont {{Pani}}}]{Brito2017}%
	\BibitemOpen
	\bibfield  {author} {\bibinfo {author} {\bibfnamefont {Richard}\ \bibnamefont
			{Brito}}, \bibinfo {author} {\bibfnamefont {S.}~\bibnamefont {{Ghosh}}},
		\bibinfo {author} {\bibfnamefont {E.}~\bibnamefont {{Barausse}}}, \bibinfo
		{author} {\bibfnamefont {E.}~\bibnamefont {{Berti}}}, \bibinfo {author}
		{\bibfnamefont {V.}~\bibnamefont {{Cardoso}}}, \bibinfo {author}
		{\bibfnamefont {I.}~\bibnamefont {{Dvorkin}}}, \bibinfo {author}
		{\bibfnamefont {A.}~\bibnamefont {{Klein}}}, \ and\ \bibinfo {author}
		{\bibfnamefont {P.}~\bibnamefont {{Pani}}},\ }\bibfield  {title} {\enquote
		{\bibinfo {title} {{Gravitational wave searches for ultralight bosons with
					LIGO and LISA}},}\ }\href {\doibase 10.1103/PhysRevD.96.064050} {\bibfield
		{journal} {\bibinfo  {journal} {Physical Review D}\ }\textbf {\bibinfo
			{volume} {96}},\ \bibinfo {pages} {064050} (\bibinfo {year}
		{2017}{\natexlab{b}})}\BibitemShut {NoStop}%
	\bibitem [{\citenamefont {Baryakhtar}\ \emph {et~al.}(2017)\citenamefont
		{Baryakhtar}, \citenamefont {Lasenby},\ and\ \citenamefont
		{Teo}}]{Baryakhtar2017}%
	\BibitemOpen
	\bibfield  {author} {\bibinfo {author} {\bibfnamefont {Masha}\ \bibnamefont
			{Baryakhtar}}, \bibinfo {author} {\bibfnamefont {Robert}\ \bibnamefont
			{Lasenby}}, \ and\ \bibinfo {author} {\bibfnamefont {Mae}\ \bibnamefont
			{Teo}},\ }\bibfield  {title} {\enquote {\bibinfo {title} {{Black hole
					superradiance signatures of ultralight vectors}},}\ }\href {\doibase
		10.1103/PhysRevD.96.035019} {\bibfield  {journal} {\bibinfo  {journal} {Phys.
				Rev. D}\ }\textbf {\bibinfo {volume} {96}},\ \bibinfo {pages} {035019}
		(\bibinfo {year} {2017})}\BibitemShut {NoStop}%
	\bibitem [{\citenamefont {Aasi}\ \emph {et~al.}(2015)\citenamefont {Aasi} \emph
		{et~al.}}]{LIGO2014}%
	\BibitemOpen
	\bibfield  {author} {\bibinfo {author} {\bibfnamefont {J.}~\bibnamefont
			{Aasi}} \emph {et~al.} (\bibinfo {collaboration} {LSC}),\ }\bibfield  {title}
	{\enquote {\bibinfo {title} {{Advanced LIGO}},}\ }\href {\doibase
		10.1088/0264-9381/32/7/074001} {\bibfield  {journal} {\bibinfo  {journal}
			{Classical and Quantum Gravity}\ }\textbf {\bibinfo {volume} {32}},\ \bibinfo
		{pages} {074001} (\bibinfo {year} {2015})}\BibitemShut {NoStop}%
	%%CITATION = ARXIV:1411.4547;%%
	\bibitem [{\citenamefont {Acernese}\ \emph {et~al.}(2015)\citenamefont
		{Acernese} \emph {et~al.}}]{Virgo2014}%
	\BibitemOpen
	\bibfield  {author} {\bibinfo {author} {\bibfnamefont {F.}~\bibnamefont
			{Acernese}} \emph {et~al.} (\bibinfo {collaboration} {Virgo}),\ }\bibfield
	{title} {\enquote {\bibinfo {title} {{Advanced Virgo: a second-generation
					interferometric gravitational wave detector}},}\ }\href {\doibase
		10.1088/0264-9381/32/2/024001} {\bibfield  {journal} {\bibinfo  {journal}
			{Classical and Quantum Gravity}\ }\textbf {\bibinfo {volume} {32}},\ \bibinfo
		{pages} {024001} (\bibinfo {year} {2015})}\BibitemShut {NoStop}%
	%%CITATION = ARXIV:1408.3978;%%
	\bibitem [{\citenamefont {Isi}\ \emph {et~al.}(2019)\citenamefont {Isi},
		\citenamefont {Sun}, \citenamefont {Brito},\ and\ \citenamefont
		{Melatos}}]{Isi2019}%
	\BibitemOpen
	\bibfield  {author} {\bibinfo {author} {\bibfnamefont {Maximiliano}\
			\bibnamefont {Isi}}, \bibinfo {author} {\bibfnamefont {Ling}\ \bibnamefont
			{Sun}}, \bibinfo {author} {\bibfnamefont {Richard}\ \bibnamefont {Brito}}, \
		and\ \bibinfo {author} {\bibfnamefont {Andrew}\ \bibnamefont {Melatos}},\
	}\bibfield  {title} {\enquote {\bibinfo {title} {Directed searches for
				gravitational waves from ultralight bosons},}\ }\href {\doibase
		10.1103/PhysRevD.99.084042} {\bibfield  {journal} {\bibinfo  {journal} {Phys.
				Rev. D}\ }\textbf {\bibinfo {volume} {99}},\ \bibinfo {pages} {084042}
		(\bibinfo {year} {2019})}\BibitemShut {NoStop}%
	\bibitem [{\citenamefont {Tsukada}\ \emph {et~al.}(2019)\citenamefont
		{Tsukada}, \citenamefont {Callister}, \citenamefont {Matas},\ and\
		\citenamefont {Meyers}}]{Tsukada2019}%
	\BibitemOpen
	\bibfield  {author} {\bibinfo {author} {\bibfnamefont {Leo}\ \bibnamefont
			{Tsukada}}, \bibinfo {author} {\bibfnamefont {Thomas}\ \bibnamefont
			{Callister}}, \bibinfo {author} {\bibfnamefont {Andrew}\ \bibnamefont
			{Matas}}, \ and\ \bibinfo {author} {\bibfnamefont {Patrick}\ \bibnamefont
			{Meyers}},\ }\bibfield  {title} {\enquote {\bibinfo {title} {First search for
				a stochastic gravitational-wave background from ultralight bosons},}\ }\href
	{\doibase 10.1103/PhysRevD.99.103015} {\bibfield  {journal} {\bibinfo
			{journal} {Phys. Rev. D}\ }\textbf {\bibinfo {volume} {99}},\ \bibinfo
		{pages} {103015} (\bibinfo {year} {2019})}\BibitemShut {NoStop}%
	\bibitem [{\citenamefont {D'Antonio}\ \emph {et~al.}(2018)\citenamefont
		{D'Antonio}, \citenamefont {Palomba}, \citenamefont {Astone}, \citenamefont
		{Frasca}, \citenamefont {Intini}, \citenamefont {La~Rosa}, \citenamefont
		{Leaci}, \citenamefont {Mastrogiovanni}, \citenamefont {Miller},
		\citenamefont {Muciaccia}, \citenamefont {Piccinni},\ and\ \citenamefont
		{Singhal}}]{DAntonio2018}%
	\BibitemOpen
	\bibfield  {author} {\bibinfo {author} {\bibfnamefont {S.}~\bibnamefont
			{D'Antonio}}, \bibinfo {author} {\bibfnamefont {C.}~\bibnamefont {Palomba}},
		\bibinfo {author} {\bibfnamefont {P.}~\bibnamefont {Astone}}, \bibinfo
		{author} {\bibfnamefont {S.}~\bibnamefont {Frasca}}, \bibinfo {author}
		{\bibfnamefont {G.}~\bibnamefont {Intini}}, \bibinfo {author} {\bibfnamefont
			{I.}~\bibnamefont {La~Rosa}}, \bibinfo {author} {\bibfnamefont
			{P.}~\bibnamefont {Leaci}}, \bibinfo {author} {\bibfnamefont
			{S.}~\bibnamefont {Mastrogiovanni}}, \bibinfo {author} {\bibfnamefont
			{A.}~\bibnamefont {Miller}}, \bibinfo {author} {\bibfnamefont
			{F.}~\bibnamefont {Muciaccia}}, \bibinfo {author} {\bibfnamefont {O.~J.}\
			\bibnamefont {Piccinni}}, \ and\ \bibinfo {author} {\bibfnamefont
			{A.}~\bibnamefont {Singhal}},\ }\bibfield  {title} {\enquote {\bibinfo
			{title} {Semicoherent analysis method to search for continuous gravitational
				waves emitted by ultralight boson clouds around spinning black holes},}\
	}\href {\doibase 10.1103/PhysRevD.98.103017} {\bibfield  {journal} {\bibinfo
			{journal} {Phys. Rev. D}\ }\textbf {\bibinfo {volume} {98}},\ \bibinfo
		{pages} {103017} (\bibinfo {year} {2018})}\BibitemShut {NoStop}%
	\bibitem [{\citenamefont {Dergachev}\ and\ \citenamefont
		{Papa}(2019)}]{Dergachev2019}%
	\BibitemOpen
	\bibfield  {author} {\bibinfo {author} {\bibfnamefont {Vladimir}\
			\bibnamefont {Dergachev}}\ and\ \bibinfo {author} {\bibfnamefont
			{Maria~Alessandra}\ \bibnamefont {Papa}},\ }\bibfield  {title} {\enquote
		{\bibinfo {title} {Sensitivity improvements in the search for periodic
				gravitational waves using o1 ligo data},}\ }\href {\doibase
		10.1103/PhysRevLett.123.101101} {\bibfield  {journal} {\bibinfo  {journal}
			{Phys. Rev. Lett.}\ }\textbf {\bibinfo {volume} {123}},\ \bibinfo {pages}
		{101101} (\bibinfo {year} {2019})}\BibitemShut {NoStop}%
	\bibitem [{\citenamefont {Palomba}\ \emph {et~al.}(2019)\citenamefont {Palomba}
		\emph {et~al.}}]{Palomba2019}%
	\BibitemOpen
	\bibfield  {author} {\bibinfo {author} {\bibfnamefont {Cristiano}\
			\bibnamefont {Palomba}} \emph {et~al.},\ }\bibfield  {title} {\enquote
		{\bibinfo {title} {{Direct constraints on ultra-light boson mass from
					searches for continuous gravitational waves}},}\ }\href {\doibase
		10.1103/PhysRevLett.123.171101} {\bibfield  {journal} {\bibinfo  {journal}
			{Phys. Rev. Lett.}\ }\textbf {\bibinfo {volume} {123}},\ \bibinfo {pages}
		{171101} (\bibinfo {year} {2019})}\BibitemShut {NoStop}%
	\bibitem [{\citenamefont {Suvorova}\ \emph {et~al.}(2016)\citenamefont
		{Suvorova}, \citenamefont {Sun}, \citenamefont {Melatos}, \citenamefont
		{Moran},\ and\ \citenamefont {Evans}}]{Suvorova2016}%
	\BibitemOpen
	\bibfield  {author} {\bibinfo {author} {\bibfnamefont {S.}~\bibnamefont
			{Suvorova}}, \bibinfo {author} {\bibfnamefont {L.}~\bibnamefont {Sun}},
		\bibinfo {author} {\bibfnamefont {A.}~\bibnamefont {Melatos}}, \bibinfo
		{author} {\bibfnamefont {W.}~\bibnamefont {Moran}}, \ and\ \bibinfo {author}
		{\bibfnamefont {Robin~J.}\ \bibnamefont {Evans}},\ }\bibfield  {title}
	{\enquote {\bibinfo {title} {{Hidden Markov model tracking of continuous
					gravitational waves from a neutron star with wandering spin}},}\ }\href
	{\doibase 10.1103/PhysRevD.93.123009} {\bibfield  {journal} {\bibinfo
			{journal} {Physical Review D}\ }\textbf {\bibinfo {volume} {93}},\ \bibinfo
		{pages} {123009} (\bibinfo {year} {2016})}\BibitemShut {NoStop}%
	\bibitem [{\citenamefont {Abbott}\ \emph
		{et~al.}(2017{\natexlab{a}})\citenamefont {Abbott} \emph
		{et~al.}}]{ScoX1ViterbiO1}%
	\BibitemOpen
	\bibfield  {author} {\bibinfo {author} {\bibfnamefont {B.~P.}\ \bibnamefont
			{Abbott}} \emph {et~al.},\ }\bibfield  {title} {\enquote {\bibinfo {title}
			{Search for gravitational waves from scorpius x-1 in the first advanced ligo
				observing run with a hidden markov model},}\ }\href {\doibase
		10.1103/PhysRevD.95.122003} {\bibfield  {journal} {\bibinfo  {journal} {Phys.
				Rev. D}\ }\textbf {\bibinfo {volume} {95}},\ \bibinfo {pages} {122003}
		(\bibinfo {year} {2017}{\natexlab{a}})}\BibitemShut {NoStop}%
	\bibitem [{\citenamefont {Sun}\ \emph {et~al.}(2018)\citenamefont {Sun},
		\citenamefont {Melatos}, \citenamefont {Suvorova}, \citenamefont {Moran},\
		and\ \citenamefont {Evans}}]{Sun2018}%
	\BibitemOpen
	\bibfield  {author} {\bibinfo {author} {\bibfnamefont {L.}~\bibnamefont
			{Sun}}, \bibinfo {author} {\bibfnamefont {A.}~\bibnamefont {Melatos}},
		\bibinfo {author} {\bibfnamefont {S.}~\bibnamefont {Suvorova}}, \bibinfo
		{author} {\bibfnamefont {W.}~\bibnamefont {Moran}}, \ and\ \bibinfo {author}
		{\bibfnamefont {R.~J.}\ \bibnamefont {Evans}},\ }\bibfield  {title} {\enquote
		{\bibinfo {title} {Hidden markov model tracking of continuous gravitational
				waves from young supernova remnants},}\ }\href {\doibase
		10.1103/PhysRevD.97.043013} {\bibfield  {journal} {\bibinfo  {journal} {Phys.
				Rev. D}\ }\textbf {\bibinfo {volume} {97}},\ \bibinfo {pages} {043013}
		(\bibinfo {year} {2018})}\BibitemShut {NoStop}%
	\bibitem [{\citenamefont {Abbott}\ \emph {et~al.}(2019)\citenamefont {Abbott}
		\emph {et~al.}}]{LVC-catalog}%
	\BibitemOpen
	\bibfield  {author} {\bibinfo {author} {\bibfnamefont {B.~P.}\ \bibnamefont
			{Abbott}} \emph {et~al.} (\bibinfo {collaboration} {LIGO Scientific
			Collaboration and Virgo Collaboration}),\ }\bibfield  {title} {\enquote
		{\bibinfo {title} {Gwtc-1: A gravitational-wave transient catalog of compact
				binary mergers observed by ligo and virgo during the first and second
				observing runs},}\ }\href {\doibase 10.1103/PhysRevX.9.031040} {\bibfield
		{journal} {\bibinfo  {journal} {Phys. Rev. X}\ }\textbf {\bibinfo {volume}
			{9}},\ \bibinfo {pages} {031040} (\bibinfo {year} {2019})}\BibitemShut
	{NoStop}%
	\bibitem [{\citenamefont {Remillard}\ and\ \citenamefont
		{McClintock}(2006)}]{Remillard:2006fc}%
	\BibitemOpen
	\bibfield  {author} {\bibinfo {author} {\bibfnamefont {Ronald~A.}\
			\bibnamefont {Remillard}}\ and\ \bibinfo {author} {\bibfnamefont
			{Jeffrey~E.}\ \bibnamefont {McClintock}},\ }\bibfield  {title} {\enquote
		{\bibinfo {title} {{X-ray Properties of Black-Hole Binaries}},}\ }\href
	{\doibase 10.1146/annurev.astro.44.051905.092532} {\bibfield  {journal}
		{\bibinfo  {journal} {Ann. Rev. Astron. Astrophys.}\ }\textbf {\bibinfo
			{volume} {44}},\ \bibinfo {pages} {49--92} (\bibinfo {year}
		{2006})}\BibitemShut {NoStop}%
	\bibitem [{\citenamefont {{Middleton}}(2016)}]{Middleton2016}%
	\BibitemOpen
	\bibfield  {author} {\bibinfo {author} {\bibfnamefont {M.}~\bibnamefont
			{{Middleton}}},\ }\bibfield  {title} {\enquote {\bibinfo {title} {{Black Hole
					Spin: Theory and Observation}},}\ }in\ \href {\doibase
		10.1007/978-3-662-52859-4_3} {\emph {\bibinfo {booktitle} {Astrophysics of
				Black Holes: From Fundamental Aspects to Latest Developments}}},\ \bibinfo
	{series} {Astrophysics and Space Science Library}, Vol.\ \bibinfo {volume}
	{440},\ \bibinfo {editor} {edited by\ \bibinfo {editor} {\bibfnamefont
			{C.}~\bibnamefont {{Bambi}}}}\ (\bibinfo {year} {2016})\ p.~\bibinfo {pages}
	{99}\BibitemShut {NoStop}%
	\bibitem [{\citenamefont {Ghosh}\ \emph {et~al.}(2019)\citenamefont {Ghosh},
		\citenamefont {Berti}, \citenamefont {Brito},\ and\ \citenamefont
		{Richartz}}]{Ghosh:2018gaw}%
	\BibitemOpen
	\bibfield  {author} {\bibinfo {author} {\bibfnamefont {Shrobana}\
			\bibnamefont {Ghosh}}, \bibinfo {author} {\bibfnamefont {Emanuele}\
			\bibnamefont {Berti}}, \bibinfo {author} {\bibfnamefont {Richard}\
			\bibnamefont {Brito}}, \ and\ \bibinfo {author} {\bibfnamefont {Mauricio}\
			\bibnamefont {Richartz}},\ }\bibfield  {title} {\enquote {\bibinfo {title}
			{{Follow-up signals from superradiant instabilities of black hole merger
					remnants}},}\ }\href {\doibase 10.1103/PhysRevD.99.104030} {\bibfield
		{journal} {\bibinfo  {journal} {Phys. Rev.}\ }\textbf {\bibinfo {volume}
			{D99}},\ \bibinfo {pages} {104030} (\bibinfo {year} {2019})}\BibitemShut
	{NoStop}%
	\bibitem [{\citenamefont {Hild}\ \emph {et~al.}(2011)\citenamefont {Hild} \emph
		{et~al.}}]{Hild:2010id}%
	\BibitemOpen
	\bibfield  {author} {\bibinfo {author} {\bibfnamefont {S.}~\bibnamefont
			{Hild}} \emph {et~al.},\ }\bibfield  {title} {\enquote {\bibinfo {title}
			{{Sensitivity Studies for Third-Generation Gravitational Wave
					Observatories}},}\ }\href {\doibase 10.1088/0264-9381/28/9/094013} {\bibfield
		{journal} {\bibinfo  {journal} {Classical and Quantum Gravity}\ }\textbf
		{\bibinfo {volume} {28}},\ \bibinfo {pages} {094013} (\bibinfo {year}
		{2011})}\BibitemShut {NoStop}%
	%%CITATION = ARXIV:1012.0908;%%
	\bibitem [{\citenamefont {Sathyaprakash}\ \emph {et~al.}(2012)\citenamefont
		{Sathyaprakash} \emph {et~al.}}]{Sathyaprakash:2012jk}%
	\BibitemOpen
	\bibfield  {author} {\bibinfo {author} {\bibfnamefont {B.}~\bibnamefont
			{Sathyaprakash}} \emph {et~al.},\ }\bibfield  {title} {\enquote {\bibinfo
			{title} {{Scientific Objectives of Einstein Telescope}},}\ }\href {\doibase
		10.1088/0264-9381/29/12/124013} {\bibfield  {journal} {\bibinfo  {journal}
			{Classical and Quantum Gravity}\ }\textbf {\bibinfo {volume} {29}},\ \bibinfo
		{pages} {124013} (\bibinfo {year} {2012})}\BibitemShut {NoStop}%
	%%CITATION = ARXIV:1206.0331;%%
	\bibitem [{\citenamefont {Punturo}\ \emph {et~al.}(2010)\citenamefont {Punturo}
		\emph {et~al.}}]{Punturo:2010zz}%
	\BibitemOpen
	\bibfield  {author} {\bibinfo {author} {\bibfnamefont {M.}~\bibnamefont
			{Punturo}} \emph {et~al.},\ }\bibfield  {title} {\enquote {\bibinfo {title}
			{{The Einstein Telescope: A third-generation gravitational wave
					observatory}},}\ }\bibfield  {booktitle} {\emph {\bibinfo {booktitle}
			{{Proceedings, 14th Workshop on Gravitational wave data analysis (GWDAW-14):
					Rome, Italy, January 26-29, 2010}}},\ }\href {\doibase
		10.1088/0264-9381/27/19/194002} {\bibfield  {journal} {\bibinfo  {journal}
			{Classical and Quantum Gravity}\ }\textbf {\bibinfo {volume} {27}},\ \bibinfo
		{pages} {194002} (\bibinfo {year} {2010})}\BibitemShut {NoStop}%
	%%CITATION = CQGRD,27,194002;%%
	\bibitem [{\citenamefont {Abbott}\ \emph
		{et~al.}(2017{\natexlab{b}})\citenamefont {Abbott} \emph
		{et~al.}}]{Abbott2017-nextGen-CE}%
	\BibitemOpen
	\bibfield  {author} {\bibinfo {author} {\bibfnamefont {Benjamin~P}\
			\bibnamefont {Abbott}} \emph {et~al.} (\bibinfo {collaboration} {LIGO
			Scientific}),\ }\bibfield  {title} {\enquote {\bibinfo {title} {{Exploring
					the Sensitivity of Next Generation Gravitational Wave Detectors}},}\ }\href
	{\doibase 10.1088/1361-6382/aa51f4} {\bibfield  {journal} {\bibinfo
			{journal} {Class. Quant. Grav.}\ }\textbf {\bibinfo {volume} {34}},\ \bibinfo
		{pages} {044001} (\bibinfo {year} {2017}{\natexlab{b}})},\ \Eprint
	{http://arxiv.org/abs/1607.08697} {arXiv:1607.08697 [astro-ph.IM]}
	\BibitemShut {NoStop}%
	%%CITATION = ARXIV:1607.08697;%%
	\bibitem [{\citenamefont {Cardoso}\ \emph {et~al.}(2018)\citenamefont
		{Cardoso}, \citenamefont {Dias}, \citenamefont {Hartnett}, \citenamefont
		{Middleton}, \citenamefont {Pani},\ and\ \citenamefont
		{Santos}}]{Cardoso2018}%
	\BibitemOpen
	\bibfield  {author} {\bibinfo {author} {\bibfnamefont {Vitor}\ \bibnamefont
			{Cardoso}}, \bibinfo {author} {\bibfnamefont {{\'{O}}scar~J.C.}\ \bibnamefont
			{Dias}}, \bibinfo {author} {\bibfnamefont {Gavin~S.}\ \bibnamefont
			{Hartnett}}, \bibinfo {author} {\bibfnamefont {Matthew}\ \bibnamefont
			{Middleton}}, \bibinfo {author} {\bibfnamefont {Paolo}\ \bibnamefont {Pani}},
		\ and\ \bibinfo {author} {\bibfnamefont {Jorge~E.}\ \bibnamefont {Santos}},\
	}\bibfield  {title} {\enquote {\bibinfo {title} {Constraining the mass of
				dark photons and axion-like particles through black-hole superradiance},}\
	}\href {\doibase 10.1088/1475-7516/2018/03/043} {\bibfield  {journal}
		{\bibinfo  {journal} {Journal of Cosmology and Astroparticle Physics}\
		}\textbf {\bibinfo {volume} {2018}},\ \bibinfo {pages} {043--043} (\bibinfo
		{year} {2018})}\BibitemShut {NoStop}%
	\bibitem [{\citenamefont {Stott}\ and\ \citenamefont
		{Marsh}(2018)}]{Stott:2018opm}%
	\BibitemOpen
	\bibfield  {author} {\bibinfo {author} {\bibfnamefont {Matthew~J.}\
			\bibnamefont {Stott}}\ and\ \bibinfo {author} {\bibfnamefont {David J.~E.}\
			\bibnamefont {Marsh}},\ }\bibfield  {title} {\enquote {\bibinfo {title}
			{{Black hole spin constraints on the mass spectrum and number of axionlike
					fields}},}\ }\href {\doibase 10.1103/PhysRevD.98.083006} {\bibfield
		{journal} {\bibinfo  {journal} {Phys. Rev.}\ }\textbf {\bibinfo {volume}
			{D98}},\ \bibinfo {pages} {083006} (\bibinfo {year} {2018})}\BibitemShut
	{NoStop}%
	\bibitem [{Note1()}]{Note1}%
	\BibitemOpen
	\bibinfo {note} {Since the QCD axion mass largely depends on $f_a$ of the
		Peccei-Quinn symmetry, e.g., $\mu \approx 6\times 10^{-10}\protect \tmspace
		+\thinmuskip {.1667em}{\protect \rm eV}(10^{16}\protect \tmspace +\thinmuskip
		{.1667em}{\protect \rm GeV}/f_a)$ \cite {Arvanitaki2010}, the BHs observable
		by aLIGO correspond to axions with $f_a$ between the GUT and Planck
		scales.}\BibitemShut {Stop}%
	\bibitem [{\citenamefont {Yoshino}\ and\ \citenamefont
		{Kodama}(2012)}]{Yoshino2012}%
	\BibitemOpen
	\bibfield  {author} {\bibinfo {author} {\bibfnamefont {Hirotaka}\
			\bibnamefont {Yoshino}}\ and\ \bibinfo {author} {\bibfnamefont {Hideo}\
			\bibnamefont {Kodama}},\ }\bibfield  {title} {\enquote {\bibinfo {title}
			{{Bosenova Collapse of Axion Cloud around a Rotating Black Hole: }},}\ }\href
	{\doibase 10.1143/PTP.128.153} {\bibfield  {journal} {\bibinfo  {journal}
			{Progress of Theoretical Physics}\ }\textbf {\bibinfo {volume} {128}},\
		\bibinfo {pages} {153--190} (\bibinfo {year} {2012})}\BibitemShut {NoStop}%
	\bibitem [{\citenamefont {Yoshino}\ and\ \citenamefont
		{Kodama}(2015{\natexlab{b}})}]{Yoshino:2015nsa}%
	\BibitemOpen
	\bibfield  {author} {\bibinfo {author} {\bibfnamefont {Hirotaka}\
			\bibnamefont {Yoshino}}\ and\ \bibinfo {author} {\bibfnamefont {Hideo}\
			\bibnamefont {Kodama}},\ }\bibfield  {title} {\enquote {\bibinfo {title}
			{{The bosenova and axiverse}},}\ }\href {\doibase
		10.1088/0264-9381/32/21/214001} {\bibfield  {journal} {\bibinfo  {journal}
			{Class. Quant. Grav.}\ }\textbf {\bibinfo {volume} {32}},\ \bibinfo {pages}
		{214001} (\bibinfo {year} {2015}{\natexlab{b}})}\BibitemShut {NoStop}%
	\bibitem [{\citenamefont {{LIGO Scientific Collaboration and Virgo
				Collaboration}}(2019)}]{GWOSC}%
	\BibitemOpen
	\bibfield  {author} {\bibinfo {author} {\bibnamefont {{LIGO Scientific
					Collaboration and Virgo Collaboration}}},\ }\href {\doibase
		10.7935/CA75-FM95} {\enquote {\bibinfo {title} {{The O2 Data Release}},}\
	}\bibinfo {howpublished}
	{\href{https://doi.org/10.7935/CA75-FM95}{https://doi.org/10.7935/CA75-FM95}}
	(\bibinfo {year} {2019})\BibitemShut {NoStop}%
	\bibitem [{\citenamefont {Vallisneri}\ \emph {et~al.}(2015)\citenamefont
		{Vallisneri}, \citenamefont {Kanner}, \citenamefont {Williams}, \citenamefont
		{Weinstein},\ and\ \citenamefont {Stephens}}]{Vallisneri:2014vxa}%
	\BibitemOpen
	\bibfield  {author} {\bibinfo {author} {\bibfnamefont {Michele}\ \bibnamefont
			{Vallisneri}}, \bibinfo {author} {\bibfnamefont {Jonah}\ \bibnamefont
			{Kanner}}, \bibinfo {author} {\bibfnamefont {Roy}\ \bibnamefont {Williams}},
		\bibinfo {author} {\bibfnamefont {Alan}\ \bibnamefont {Weinstein}}, \ and\
		\bibinfo {author} {\bibfnamefont {Branson}\ \bibnamefont {Stephens}},\
	}\bibfield  {title} {\enquote {\bibinfo {title} {{The LIGO Open Science
					Center}},}\ }\bibfield  {booktitle} {\emph {\bibinfo {booktitle}
			{{Proceedings, 10th International LISA Symposium: Gainesville, Florida, USA,
					May 18-23, 2014}}},\ }\href {\doibase 10.1088/1742-6596/610/1/012021}
	{\bibfield  {journal} {\bibinfo  {journal} {J. Phys. Conf. Ser.}\ }\textbf
		{\bibinfo {volume} {610}},\ \bibinfo {pages} {012021} (\bibinfo {year}
		{2015})},\ \Eprint {http://arxiv.org/abs/1410.4839} {arXiv:1410.4839 [gr-qc]}
	\BibitemShut {NoStop}%
	%%CITATION = ARXIV:1410.4839;%%
	\bibitem [{\citenamefont {Reynolds}(2014)}]{Reynolds:2013qqa}%
	\BibitemOpen
	\bibfield  {author} {\bibinfo {author} {\bibfnamefont {Christopher~S.}\
			\bibnamefont {Reynolds}},\ }\bibfield  {title} {\enquote {\bibinfo {title}
			{{Measuring Black Hole Spin using X-ray Reflection Spectroscopy}},}\ }\href
	{\doibase 10.1007/s11214-013-0006-6} {\bibfield  {journal} {\bibinfo
			{journal} {Space Sci. Rev.}\ }\textbf {\bibinfo {volume} {183}},\ \bibinfo
		{pages} {277--294} (\bibinfo {year} {2014})}\BibitemShut {NoStop}%
	\bibitem [{\citenamefont {McClintock}\ \emph {et~al.}(2014)\citenamefont
		{McClintock}, \citenamefont {Narayan},\ and\ \citenamefont
		{Steiner}}]{McClintock:2013vwa}%
	\BibitemOpen
	\bibfield  {author} {\bibinfo {author} {\bibfnamefont {Jeffrey~E.}\
			\bibnamefont {McClintock}}, \bibinfo {author} {\bibfnamefont {Ramesh}\
			\bibnamefont {Narayan}}, \ and\ \bibinfo {author} {\bibfnamefont {James~F.}\
			\bibnamefont {Steiner}},\ }\bibfield  {title} {\enquote {\bibinfo {title}
			{{Black Hole Spin via Continuum Fitting and the Role of Spin in Powering
					Transient Jets}},}\ }\href {\doibase 10.1007/s11214-013-0003-9} {\bibfield
		{journal} {\bibinfo  {journal} {Space Sci. Rev.}\ }\textbf {\bibinfo {volume}
			{183}},\ \bibinfo {pages} {295--322} (\bibinfo {year} {2014})}\BibitemShut
	{NoStop}%
	\bibitem [{\citenamefont {Gou}\ \emph {et~al.}(2011)\citenamefont {Gou},
		\citenamefont {McClintock}, \citenamefont {Reid}, \citenamefont {Orosz},
		\citenamefont {Steiner}, \citenamefont {Narayan}, \citenamefont {Xiang},
		\citenamefont {Remillard}, \citenamefont {Arnaud},\ and\ \citenamefont
		{Davis}}]{Gou2011}%
	\BibitemOpen
	\bibfield  {author} {\bibinfo {author} {\bibfnamefont {Lijun}\ \bibnamefont
			{Gou}}, \bibinfo {author} {\bibfnamefont {Jeffrey~E.}\ \bibnamefont
			{McClintock}}, \bibinfo {author} {\bibfnamefont {Mark~J.}\ \bibnamefont
			{Reid}}, \bibinfo {author} {\bibfnamefont {Jerome~A.}\ \bibnamefont {Orosz}},
		\bibinfo {author} {\bibfnamefont {James~F.}\ \bibnamefont {Steiner}},
		\bibinfo {author} {\bibfnamefont {Ramesh}\ \bibnamefont {Narayan}}, \bibinfo
		{author} {\bibfnamefont {Jingen}\ \bibnamefont {Xiang}}, \bibinfo {author}
		{\bibfnamefont {Ronald~A.}\ \bibnamefont {Remillard}}, \bibinfo {author}
		{\bibfnamefont {Keith~A.}\ \bibnamefont {Arnaud}}, \ and\ \bibinfo {author}
		{\bibfnamefont {Shane~W.}\ \bibnamefont {Davis}},\ }\bibfield  {title}
	{\enquote {\bibinfo {title} {{THE} {EXTREME} {SPIN} {OF} {THE} {BLACK} {HOLE}
				{IN} {CYGNUS} x-1},}\ }\href {\doibase 10.1088/0004-637x/742/2/85} {\bibfield
		{journal} {\bibinfo  {journal} {The Astrophysical Journal}\ }\textbf
		{\bibinfo {volume} {742}},\ \bibinfo {pages} {85} (\bibinfo {year}
		{2011})}\BibitemShut {NoStop}%
	\bibitem [{\citenamefont {Wong}\ \emph {et~al.}(2012)\citenamefont {Wong},
		\citenamefont {Valsecchi}, \citenamefont {Fragos},\ and\ \citenamefont
		{Kalogera}}]{Wong2012}%
	\BibitemOpen
	\bibfield  {author} {\bibinfo {author} {\bibfnamefont {Tsing-Wai}\
			\bibnamefont {Wong}}, \bibinfo {author} {\bibfnamefont {Francesca}\
			\bibnamefont {Valsecchi}}, \bibinfo {author} {\bibfnamefont {Tassos}\
			\bibnamefont {Fragos}}, \ and\ \bibinfo {author} {\bibfnamefont {Vassiliki}\
			\bibnamefont {Kalogera}},\ }\bibfield  {title} {\enquote {\bibinfo {title}
			{{UNDERSTANDING} {COMPACT} {OBJECT} {FORMATION} {AND} {NATAL} {KICKS}. {III}.
				{THE} {CASE} {OF} {CYGNUS} x-1},}\ }\href {\doibase
		10.1088/0004-637x/747/2/111} {\bibfield  {journal} {\bibinfo  {journal} {The
				Astrophysical Journal}\ }\textbf {\bibinfo {volume} {747}},\ \bibinfo {pages}
		{111} (\bibinfo {year} {2012})}\BibitemShut {NoStop}%
	\bibitem [{\citenamefont {Axelsson}\ \emph {et~al.}(2011)\citenamefont
		{Axelsson}, \citenamefont {Church}, \citenamefont {Davies}, \citenamefont
		{Levan},\ and\ \citenamefont {Ryde}}]{Axelsson2011}%
	\BibitemOpen
	\bibfield  {author} {\bibinfo {author} {\bibfnamefont {Magnus}\ \bibnamefont
			{Axelsson}}, \bibinfo {author} {\bibfnamefont {Ross~P.}\ \bibnamefont
			{Church}}, \bibinfo {author} {\bibfnamefont {Melvyn~B.}\ \bibnamefont
			{Davies}}, \bibinfo {author} {\bibfnamefont {Andrew~J.}\ \bibnamefont
			{Levan}}, \ and\ \bibinfo {author} {\bibfnamefont {Felix}\ \bibnamefont
			{Ryde}},\ }\bibfield  {title} {\enquote {\bibinfo {title} {{On the origin of
					black hole spin in high-mass black hole binaries: Cygnus X-1}},}\ }\href
	{\doibase 10.1111/j.1365-2966.2010.18050.x} {\bibfield  {journal} {\bibinfo
			{journal} {Monthly Notices of the Royal Astronomical Society}\ }\textbf
		{\bibinfo {volume} {412}},\ \bibinfo {pages} {2260--2264} (\bibinfo {year}
		{2011})}\BibitemShut {NoStop}%
	\bibitem [{\citenamefont {Walton}\ \emph {et~al.}(2016)\citenamefont {Walton},
		\citenamefont {Tomsick}, \citenamefont {Madsen}, \citenamefont {Grinberg},
		\citenamefont {Barret}, \citenamefont {Boggs}, \citenamefont {Christensen},
		\citenamefont {Clavel}, \citenamefont {Craig}, \citenamefont {Fabian},
		\citenamefont {Fuerst}, \citenamefont {Hailey}, \citenamefont {Harrison},
		\citenamefont {Miller}, \citenamefont {Parker}, \citenamefont {Rahoui},
		\citenamefont {Stern}, \citenamefont {Tao}, \citenamefont {Wilms},\ and\
		\citenamefont {Zhang}}]{Walton2016}%
	\BibitemOpen
	\bibfield  {author} {\bibinfo {author} {\bibfnamefont {D.~J.}\ \bibnamefont
			{Walton}}, \bibinfo {author} {\bibfnamefont {J.~A.}\ \bibnamefont {Tomsick}},
		\bibinfo {author} {\bibfnamefont {K.~K.}\ \bibnamefont {Madsen}}, \bibinfo
		{author} {\bibfnamefont {V.}~\bibnamefont {Grinberg}}, \bibinfo {author}
		{\bibfnamefont {D.}~\bibnamefont {Barret}}, \bibinfo {author} {\bibfnamefont
			{S.~E.}\ \bibnamefont {Boggs}}, \bibinfo {author} {\bibfnamefont {F.~E.}\
			\bibnamefont {Christensen}}, \bibinfo {author} {\bibfnamefont
			{M.}~\bibnamefont {Clavel}}, \bibinfo {author} {\bibfnamefont {W.~W.}\
			\bibnamefont {Craig}}, \bibinfo {author} {\bibfnamefont {A.~C.}\ \bibnamefont
			{Fabian}}, \bibinfo {author} {\bibfnamefont {F.}~\bibnamefont {Fuerst}},
		\bibinfo {author} {\bibfnamefont {C.~J.}\ \bibnamefont {Hailey}}, \bibinfo
		{author} {\bibfnamefont {F.~A.}\ \bibnamefont {Harrison}}, \bibinfo {author}
		{\bibfnamefont {J.~M.}\ \bibnamefont {Miller}}, \bibinfo {author}
		{\bibfnamefont {M.~L.}\ \bibnamefont {Parker}}, \bibinfo {author}
		{\bibfnamefont {F.}~\bibnamefont {Rahoui}}, \bibinfo {author} {\bibfnamefont
			{D.}~\bibnamefont {Stern}}, \bibinfo {author} {\bibfnamefont
			{L.}~\bibnamefont {Tao}}, \bibinfo {author} {\bibfnamefont {J.}~\bibnamefont
			{Wilms}}, \ and\ \bibinfo {author} {\bibfnamefont {W.}~\bibnamefont
			{Zhang}},\ }\bibfield  {title} {\enquote {\bibinfo {title} {{THE} {SOFT}
				{STATE} {OF} {CYGNUS} x-1 {OBSERVED} {WITHNuSTAR}: A {VARIABLE} {CORONA}
				{AND} a {STABLE} {INNER} {DISK}},}\ }\href {\doibase
		10.3847/0004-637x/826/1/87} {\bibfield  {journal} {\bibinfo  {journal} {The
				Astrophysical Journal}\ }\textbf {\bibinfo {volume} {826}},\ \bibinfo {pages}
		{87} (\bibinfo {year} {2016})}\BibitemShut {NoStop}%
	\bibitem [{\citenamefont {Miller}\ \emph {et~al.}(2009)\citenamefont {Miller},
		\citenamefont {Reynolds}, \citenamefont {Fabian}, \citenamefont {Miniutti},\
		and\ \citenamefont {Gallo}}]{Miller2009}%
	\BibitemOpen
	\bibfield  {author} {\bibinfo {author} {\bibfnamefont {J.~M.}\ \bibnamefont
			{Miller}}, \bibinfo {author} {\bibfnamefont {C.~S.}\ \bibnamefont
			{Reynolds}}, \bibinfo {author} {\bibfnamefont {A.~C.}\ \bibnamefont
			{Fabian}}, \bibinfo {author} {\bibfnamefont {G.}~\bibnamefont {Miniutti}}, \
		and\ \bibinfo {author} {\bibfnamefont {L.~C.}\ \bibnamefont {Gallo}},\
	}\bibfield  {title} {\enquote {\bibinfo {title} {{STELLAR}-{MASS} {BLACK}
				{HOLE} {SPIN} {CONSTRAINTS} {FROM} {DISK} {REFLECTION} {AND} {CONTINUUM}
				{MODELING}},}\ }\href {\doibase 10.1088/0004-637x/697/1/900} {\bibfield
		{journal} {\bibinfo  {journal} {The Astrophysical Journal}\ }\textbf
		{\bibinfo {volume} {697}},\ \bibinfo {pages} {900--912} (\bibinfo {year}
		{2009})}\BibitemShut {NoStop}%
	\bibitem [{\citenamefont {Kawano}\ \emph {et~al.}(2017)\citenamefont {Kawano},
		\citenamefont {Done}, \citenamefont {Yamada}, \citenamefont {Takahashi},
		\citenamefont {Axelsson},\ and\ \citenamefont {Fukazawa}}]{Kawano2017}%
	\BibitemOpen
	\bibfield  {author} {\bibinfo {author} {\bibfnamefont {Takafumi}\
			\bibnamefont {Kawano}}, \bibinfo {author} {\bibfnamefont {Chris}\
			\bibnamefont {Done}}, \bibinfo {author} {\bibfnamefont {Shin’ya}\
			\bibnamefont {Yamada}}, \bibinfo {author} {\bibfnamefont {Hiromitsu}\
			\bibnamefont {Takahashi}}, \bibinfo {author} {\bibfnamefont {Magnus}\
			\bibnamefont {Axelsson}}, \ and\ \bibinfo {author} {\bibfnamefont {Yasushi}\
			\bibnamefont {Fukazawa}},\ }\bibfield  {title} {\enquote {\bibinfo {title}
			{{Black hole spin of Cygnus X-1 determined from the softest state ever
					observed}},}\ }\href {\doibase 10.1093/pasj/psx009} {\bibfield  {journal}
		{\bibinfo  {journal} {Publications of the Astronomical Society of Japan}\
		}\textbf {\bibinfo {volume} {69}} (\bibinfo {year} {2017}),\
		10.1093/pasj/psx009},\ \bibinfo {note} {36}\BibitemShut {NoStop}%
	\bibitem [{\citenamefont {Krawczynski}(2018)}]{Krawczynski2018}%
	\BibitemOpen
	\bibfield  {author} {\bibinfo {author} {\bibfnamefont {Henric}\ \bibnamefont
			{Krawczynski}},\ }\bibfield  {title} {\enquote {\bibinfo {title}
			{Difficulties of quantitative tests of the kerr-hypothesis with x-ray
				observations of mass accreting black holes},}\ }\href {\doibase
		10.1007/s10714-018-2419-8} {\bibfield  {journal} {\bibinfo  {journal}
			{General Relativity and Gravitation}\ }\textbf {\bibinfo {volume} {50}},\
		\bibinfo {pages} {100} (\bibinfo {year} {2018})}\BibitemShut {NoStop}%
	\bibitem [{Note2()}]{Note2}%
	\BibitemOpen
	\bibinfo {note} {An improved HMM method, which tracks the binary orbital
		phase and sums the signal power in orbital sidebands coherently, proves to be
		more sensitive \cite {Suvorova2017}. However, the improved method depends on
		the measurement of the time of passage through the ascending node, which is
		not available for Cyg X-1. Hence we do not apply the orbital phase tracking
		in this analysis.}\BibitemShut {Stop}%
	\bibitem [{\citenamefont {Viterbi}(1967)}]{Viterbi1967}%
	\BibitemOpen
	\bibfield  {author} {\bibinfo {author} {\bibfnamefont {A.}~\bibnamefont
			{Viterbi}},\ }\bibfield  {title} {\enquote {\bibinfo {title} {{Error bounds
					for convolutional codes and an asymptotically optimum decoding algorithm}},}\
	}\href {\doibase 10.1109/TIT.1967.1054010} {\bibfield  {journal} {\bibinfo
			{journal} {IEEE Transactions on Information Theory}\ }\textbf {\bibinfo
			{volume} {13}},\ \bibinfo {pages} {260--269} (\bibinfo {year}
		{1967})}\BibitemShut {NoStop}%
	\bibitem [{\citenamefont {Orosz}\ \emph {et~al.}(2011)\citenamefont {Orosz},
		\citenamefont {McClintock}, \citenamefont {Aufdenberg}, \citenamefont
		{Remillard}, \citenamefont {Reid}, \citenamefont {Narayan},\ and\
		\citenamefont {Gou}}]{Orosz2011}%
	\BibitemOpen
	\bibfield  {author} {\bibinfo {author} {\bibfnamefont {Jerome~A.}\
			\bibnamefont {Orosz}}, \bibinfo {author} {\bibfnamefont {Jeffrey~E.}\
			\bibnamefont {McClintock}}, \bibinfo {author} {\bibfnamefont {Jason~P.}\
			\bibnamefont {Aufdenberg}}, \bibinfo {author} {\bibfnamefont {Ronald~A.}\
			\bibnamefont {Remillard}}, \bibinfo {author} {\bibfnamefont {Mark~J.}\
			\bibnamefont {Reid}}, \bibinfo {author} {\bibfnamefont {Ramesh}\ \bibnamefont
			{Narayan}}, \ and\ \bibinfo {author} {\bibfnamefont {Lijun}\ \bibnamefont
			{Gou}},\ }\bibfield  {title} {\enquote {\bibinfo {title} {{The Mass of the
					Black Hole in Cygnus X-1}},}\ }\href {\doibase 10.1088/0004-637X/742/2/84}
	{\bibfield  {journal} {\bibinfo  {journal} {Astrophys. J.}\ }\textbf
		{\bibinfo {volume} {742}},\ \bibinfo {pages} {84} (\bibinfo {year}
		{2011})}\BibitemShut {NoStop}%
	\bibitem [{\citenamefont {Casares}\ and\ \citenamefont
		{Jonker}(2014)}]{Casares2014}%
	\BibitemOpen
	\bibfield  {author} {\bibinfo {author} {\bibfnamefont {J.}~\bibnamefont
			{Casares}}\ and\ \bibinfo {author} {\bibfnamefont {P.~G.}\ \bibnamefont
			{Jonker}},\ }\bibfield  {title} {\enquote {\bibinfo {title} {Mass
				measurements of stellar and intermediate-mass black holes},}\ }\href
	{\doibase 10.1007/s11214-013-0030-6} {\bibfield  {journal} {\bibinfo
			{journal} {Space Science Reviews}\ }\textbf {\bibinfo {volume} {183}},\
		\bibinfo {pages} {223--252} (\bibinfo {year} {2014})}\BibitemShut {NoStop}%
	\bibitem [{\citenamefont {{Reid}}\ \emph {et~al.}(2011)\citenamefont {{Reid}},
		\citenamefont {{McClintock}}, \citenamefont {{Narayan}}, \citenamefont
		{{Gou}}, \citenamefont {{Remillard}},\ and\ \citenamefont
		{{Orosz}}}]{Reid2011}%
	\BibitemOpen
	\bibfield  {author} {\bibinfo {author} {\bibfnamefont {M.~J.}\ \bibnamefont
			{{Reid}}}, \bibinfo {author} {\bibfnamefont {J.~E.}\ \bibnamefont
			{{McClintock}}}, \bibinfo {author} {\bibfnamefont {R.}~\bibnamefont
			{{Narayan}}}, \bibinfo {author} {\bibfnamefont {L.}~\bibnamefont {{Gou}}},
		\bibinfo {author} {\bibfnamefont {R.~A.}\ \bibnamefont {{Remillard}}}, \ and\
		\bibinfo {author} {\bibfnamefont {J.~A.}\ \bibnamefont {{Orosz}}},\
	}\bibfield  {title} {\enquote {\bibinfo {title} {{The Trigonometric Parallax
					of Cygnus X-1}},}\ }\href {\doibase 10.1088/0004-637X/742/2/83} {\bibfield
		{journal} {\bibinfo  {journal} {\apj}\ }\textbf {\bibinfo {volume} {742}},\
		\bibinfo {eid} {83} (\bibinfo {year} {2011})}\BibitemShut {NoStop}%
	\bibitem [{Note3()}]{Note3}%
	\BibitemOpen
	\bibinfo {note} {Note that throughout this paper, the GW strain $h_0$ differs
		from the numerically estimated strain in Eqn.~(28) of Ref.~\cite {Isi2019}
		due to different conventions. The strain in Eqn.~(28) of \cite {Isi2019}
		needs to be multiplied by a factor of $\protect \sqrt {5/(4\pi )}$ to be
		directly comparable to the $h_0$ in this paper \cite
		{Isi2019-erratum,Sun2020-erratum}.}\BibitemShut {Stop}%
	\bibitem [{Note4()}]{Note4}%
	\BibitemOpen
	\bibinfo {note} {The data collected in December 2016 is not used, because
		there is an end-of-year break starting on 22 December 2016 and the data
		quality before that (at the beginning of O2) is not optimal.}\BibitemShut
	{Stop}%
	\bibitem [{Note5()}]{Note5}%
	\BibitemOpen
	\bibinfo {note} {Summing the orbital sideband powers incoherently leads to
		the sensitivity loss scaling as $f^{1/4}$ \cite {Suvorova2016}.}\BibitemShut
	{Stop}%
	\bibitem [{Note6()}]{Note6}%
	\BibitemOpen
	\bibinfo {note} {Here the effective $S_h(f)$ is calculated from the harmonic
		mean of the two detectors over all the 30-min short Fourier transforms
		collected from GPS time 1180310418 to 1187733592.}\BibitemShut {Stop}%
	\bibitem [{Note7()}]{Note7}%
	\BibitemOpen
	\bibinfo {note} {The authors thank Ilya Mandel for helpful input regarding
		the potentially younger age of the BH.}\BibitemShut {Stop}%
	\bibitem [{\citenamefont {Russell}\ \emph {et~al.}(2007)\citenamefont
		{Russell}, \citenamefont {Fender}, \citenamefont {Gallo},\ and\ \citenamefont
		{Kaiser}}]{Russell2007}%
	\BibitemOpen
	\bibfield  {author} {\bibinfo {author} {\bibfnamefont {D.~M.}\ \bibnamefont
			{Russell}}, \bibinfo {author} {\bibfnamefont {R.~P.}\ \bibnamefont {Fender}},
		\bibinfo {author} {\bibfnamefont {E.}~\bibnamefont {Gallo}}, \ and\ \bibinfo
		{author} {\bibfnamefont {C.~R.}\ \bibnamefont {Kaiser}},\ }\bibfield  {title}
	{\enquote {\bibinfo {title} {{The jet-powered optical nebula of Cygnus
					X–1}},}\ }\href {\doibase 10.1111/j.1365-2966.2007.11539.x} {\bibfield
		{journal} {\bibinfo  {journal} {Monthly Notices of the Royal Astronomical
				Society}\ }\textbf {\bibinfo {volume} {376}},\ \bibinfo {pages} {1341--1349}
		(\bibinfo {year} {2007})}\BibitemShut {NoStop}%
	\bibitem [{\citenamefont {Sun}\ \emph {et~al.}(2020)\citenamefont {Sun},
		\citenamefont {Brito},\ and\ \citenamefont {Isi}}]{Sun2020-erratum}%
	\BibitemOpen
	\bibfield  {author} {\bibinfo {author} {\bibfnamefont {Ling}\ \bibnamefont
			{Sun}}, \bibinfo {author} {\bibfnamefont {Richard}\ \bibnamefont {Brito}}, \
		and\ \bibinfo {author} {\bibfnamefont {Maximiliano}\ \bibnamefont {Isi}},\
	}\bibfield  {title} {\enquote {\bibinfo {title} {Erratum: Search for
				ultralight bosons in cygnus x-1 with advanced ligo [phys. rev. d 101, 063020
				(2020)]},}\ }\href {\doibase 10.1103/PhysRevD.102.089902} {\bibfield
		{journal} {\bibinfo  {journal} {Phys. Rev. D}\ }\textbf {\bibinfo {volume}
			{102}},\ \bibinfo {pages} {089902} (\bibinfo {year} {2020})}\BibitemShut
	{NoStop}%
	\bibitem [{\citenamefont {Quinn}\ and\ \citenamefont
		{Hannan}(2001)}]{Quinn2001}%
	\BibitemOpen
	\bibfield  {author} {\bibinfo {author} {\bibfnamefont {B.~G.}\ \bibnamefont
			{Quinn}}\ and\ \bibinfo {author} {\bibfnamefont {E.~J.}\ \bibnamefont
			{Hannan}},\ }\href@noop {} {\emph {\bibinfo {title} {{The Estimation and
					Tracking of Frequency}}}}\ (\bibinfo  {publisher} {Cambridge University
		Press},\ \bibinfo {year} {2001})\ p.\ \bibinfo {pages} {266}\BibitemShut
	{NoStop}%
	\bibitem [{\citenamefont {Jaranowski}\ \emph {et~al.}(1998)\citenamefont
		{Jaranowski}, \citenamefont {Kr{\'{o}}lak},\ and\ \citenamefont
		{Schutz}}]{Jaranowski1998}%
	\BibitemOpen
	\bibfield  {author} {\bibinfo {author} {\bibfnamefont {Piotr}\ \bibnamefont
			{Jaranowski}}, \bibinfo {author} {\bibfnamefont {Andrzej}\ \bibnamefont
			{Kr{\'{o}}lak}}, \ and\ \bibinfo {author} {\bibfnamefont {Bernard~F.}\
			\bibnamefont {Schutz}},\ }\bibfield  {title} {\enquote {\bibinfo {title}
			{{Data analysis of gravitational-wave signals from spinning neutron stars:
					The signal and its detection}},}\ }\href {\doibase
		10.1103/PhysRevD.58.063001} {\bibfield  {journal} {\bibinfo  {journal}
			{Physical Review D}\ }\textbf {\bibinfo {volume} {58}},\ \bibinfo {pages}
		{063001} (\bibinfo {year} {1998})}\BibitemShut {NoStop}%
	\bibitem [{\citenamefont {Prix}(Apr 2019)}]{F-stat2019}%
	\BibitemOpen
	\bibfield  {author} {\bibinfo {author} {\bibfnamefont {Reinhard}\
			\bibnamefont {Prix}},\ }\bibfield  {title} {\enquote {\bibinfo {title} {{The
					$\mathcal{F}$-statistic and its implementation in ComputeFstatistic v2}},}\
	}\href {https://dcc.ligo.org/LIGO-T0900149/public} {\bibfield  {journal}
		{\bibinfo  {journal} {LIGO Document T0900149}\ } (\bibinfo {year} {Apr
			2019})}\BibitemShut {NoStop}%
	\bibitem [{\citenamefont {Abramowitz}\ and\ \citenamefont
		{Stegun}(1964)}]{Abramowitz1964}%
	\BibitemOpen
	\bibfield  {author} {\bibinfo {author} {\bibfnamefont {Milton}\ \bibnamefont
			{Abramowitz}}\ and\ \bibinfo {author} {\bibfnamefont {Irene~A.}\ \bibnamefont
			{Stegun}},\ }\href@noop {} {\emph {\bibinfo {title} {{Handbook of
					Mathematical Functions: With Formulas, Graphs, and Mathematical Tables}}}}\
	(\bibinfo  {publisher} {Courier Corporation},\ \bibinfo {year}
	{1964})\BibitemShut {NoStop}%
	\bibitem [{\citenamefont {Suvorova}\ \emph {et~al.}(2017)\citenamefont
		{Suvorova}, \citenamefont {Clearwater}, \citenamefont {Melatos},
		\citenamefont {Sun}, \citenamefont {Moran},\ and\ \citenamefont
		{Evans}}]{Suvorova2017}%
	\BibitemOpen
	\bibfield  {author} {\bibinfo {author} {\bibfnamefont {S.}~\bibnamefont
			{Suvorova}}, \bibinfo {author} {\bibfnamefont {P.}~\bibnamefont
			{Clearwater}}, \bibinfo {author} {\bibfnamefont {A.}~\bibnamefont {Melatos}},
		\bibinfo {author} {\bibfnamefont {L.}~\bibnamefont {Sun}}, \bibinfo {author}
		{\bibfnamefont {W.}~\bibnamefont {Moran}}, \ and\ \bibinfo {author}
		{\bibfnamefont {R.~J.}\ \bibnamefont {Evans}},\ }\bibfield  {title} {\enquote
		{\bibinfo {title} {{Hidden Markov model tracking of continuous gravitational
					waves from a binary neutron star with wandering spin. II. Binary orbital
					phase tracking}},}\ }\href {\doibase 10.1103/PhysRevD.96.102006} {\bibfield
		{journal} {\bibinfo  {journal} {Phys. Rev.}\ }\textbf {\bibinfo {volume}
			{D96}},\ \bibinfo {pages} {102006} (\bibinfo {year} {2017})}\BibitemShut
	{NoStop}%
	\bibitem [{\citenamefont {Isi}\ \emph {et~al.}(2020)\citenamefont {Isi},
		\citenamefont {Sun}, \citenamefont {Brito},\ and\ \citenamefont
		{Melatos}}]{Isi2019-erratum}%
	\BibitemOpen
	\bibfield  {author} {\bibinfo {author} {\bibfnamefont {Maximiliano}\
			\bibnamefont {Isi}}, \bibinfo {author} {\bibfnamefont {Ling}\ \bibnamefont
			{Sun}}, \bibinfo {author} {\bibfnamefont {Richard}\ \bibnamefont {Brito}}, \
		and\ \bibinfo {author} {\bibfnamefont {Andrew}\ \bibnamefont {Melatos}},\
	}\bibfield  {title} {\enquote {\bibinfo {title} {Erratum: Directed searches
				for gravitational waves from ultralight bosons [phys. rev. d 99, 084042
				(2019)]},}\ }\href {\doibase 10.1103/PhysRevD.102.049901} {\bibfield
		{journal} {\bibinfo  {journal} {Phys. Rev. D}\ }\textbf {\bibinfo {volume}
			{102}},\ \bibinfo {pages} {049901} (\bibinfo {year} {2020})}\BibitemShut
	{NoStop}%
\end{thebibliography}
\end{document}